\documentclass[preprint,12pt]{revtex4-1}
\usepackage{graphicx}
\usepackage{dcolumn}
\usepackage{subcaption}
\usepackage[export]{adjustbox}
\usepackage{wrapfig}
\usepackage{bm}
\usepackage[utf8]{inputenc}
\usepackage[T1]{fontenc}
\usepackage[font={small}]{caption}
\usepackage{mathptmx}
\usepackage{multirow}
\usepackage{booktabs}
\draft 

\begin{document}


\title{Resonant photoproduction of ultrarelativistic electron-positron pairs on a nucleus in strong monochromatic light field}

\author{Sergei P. Roshchupkin}
\email[]{serg9rsp@gmail.com}
\affiliation{Peter the Great St.Petersburg Polytechnic University, Saint-Petersburg, Russia}

\author{Nikita R. Larin}
\email[]{larin_nr@spbstu.ru}
\affiliation{Peter the Great St.Petersburg Polytechnic University, Saint-Petersburg, Russia}

\author{Victor V. Dubov}
\email[]{dubov@spbstu.ru}
\affiliation{Peter the Great St.Petersburg Polytechnic University, Saint-Petersburg, Russia}


\begin{abstract}
For complete development of quantum electrodynamics in the presence of a strong external field, the proper understanding of resonant processes and all their peculiarities is essential. We present our attempt to analytically investigate the resonant case of laser-assisted electron-positron pair photoproduction on a nucleus. Due to the presence of external field, the intermediate virtual particle may become real, herewith the second order process in the fine structure constant effectively reduces into the two successive first order processes. All inherent kinematics features were discussed in details and the resonant differential cross section was obtained. We established that the resonant energies of produced particles ambiguously depend on the positron (channel A) or electron (channel B) outgoing angle, and the certain minimal amount of absorbed wave photons are required for resonance to happen. Furthermore, the resonant cross section significantly exceeds the corresponding one in the absence of the external field within the particular kinematic regions and consequently, the considered process can be used qua a marker for probing theoretical predictions of quantum electrodynamics with strong background field.
\end{abstract}


\maketitle

\section{Introduction}

Nowadays, the nonlinear phenomena of quantum electrodynamics (QED) within the strong electromagnetic fields attract enormous interest \cite{PiazzaRev,gonoskov2021charged,Blackburn2020,SALAMIN200641,Ehlotzky_2009,Cajiao_V_lez_2019,Ritus,monmon,monpul} due to the development of the contemporary high-intensity laser radiation facilities \cite{danson,Yoon:21,turcu,papadopoulos,bromage,ROSSBACH20191,refId0} and high-energy particles sources \cite{Gonoskov,magnusson,Zhu_2018}. Amongst of such phenomena, the resonant behavior of the second order processes in the fine structure constant \cite{Oleinik1,Oleinik2,Kra,Bos_1979,intech,Res_rev_rsp}. The feature of these processes is that the intermediate virtual particle can possibly become a real one, and by virtue of it the initial process of the second order effectively splits into the two successive first order processes. Wherein, the resonant differential cross section may significantly surpass the corresponding non-resonant one within the certain kinematic region. Therefore, it makes resonant processes potential candidates to become markers for probing the predictions of QED in the presence of a strong external field. 

The conversion of electromagnetic radiation into the matter is one of the most intriguing phenomena since the dawn of quantum field theory. There are diverse scenarios of the electron-positron pair productions in nature \cite{RUFFINI20101,FradGit}, amid them the famous Bethe-Haitler (BH) \cite{bethe34} and Breit-Wheeler (BW) \cite{BW} processes. In turn, the former may be modified by the presence of strong external field, and one refers to it as the laser-assisted BH process. In the present paper, we are concerned with the resonant case of this process. Notwithstanding the large number of fruitful investigations devoted to the laser-assisted PPP on a nucleus \cite{L_tstedt_2009,PiazzaRelNuc,AUGUSTIN2014114,Muller,Recoil,Hafizi}, the complete description of this problem, especially the resonant situation, is hitherto far away from the completeness. We want explicitly to highlight previous works, where attention was paid to the resonant laser-assisted BH process for the case of weak monochromatic \cite{Larin,LarinModern} and pulsed \cite{LarinLasPhys,LarinUniverse} plane wave field. Within the current research, our analytical investigation is extended to the case of strong external field.

The inherent feature about the processes within external electromagnetic wave field is that there are two characteristic parameters that govern their behavior. The first is a classical relativistic invariant parameter, which defines the interactions of fermions with background plane wave field:
 \begin{eqnarray}
\eta  = \frac{{eF\mathchar'26\mkern-10mu\lambda  }}{{m{c^2}}},
\label{1}
\end{eqnarray}
which numerically equals to the ratio of the field work at a wavelength to the electron rest energy ($e$ and $m$ are the charge and the electron mass, $F$ and   $\;\mathchar'26\mkern-10mu\lambda  = {c \mathord{\left/
 {\vphantom {c \omega }} \right.
 \kern-\nulldelimiterspace} \omega }$ are the field strength and wavelength, $\omega$ is a wave frequency). The second is a quantum multiphoton parameter appears when particles interact with the Coulomb center within the plane electromagnetic wave \cite{Bunkin_Fedorov}: 
\begin{eqnarray}
\gamma  = \eta \frac{{mvc}}{{\hbar \omega }}
\label{2}
\end{eqnarray}
Herein $v$  is the particle velocity, $c$  is the speed of light. However, this parameter (\ref{2}) plays an essential role only for the case, when particles are scattered on a large angle by the Coulomb potential. Otherwise, when the scattering angle is small, this parameter does not appear \cite{Lebed__2016}. Thus, the main parameter that determines the multiphoton processes is the classical relativistic parameter (\ref{1}). Henceforth, we will employ the relativistic system of units $\hbar  = c = 1$.

\section{The amplitude of the process}

In order to deduce analytical expressions for the resonant differential cross section we adhere to the model of infinitely spatially and temporally extended electromagnetic wave with circular polarization, which propagates along the $z$ axes. Let us choose the corresponding four-potential in the following form:
\begin{eqnarray}
A\left( \phi  \right) = \frac{F}{\omega }\left( {{e_x}\cos \phi  + \delta {e_y}\sin \phi } \right),\quad \phi  = kx = \omega \left( {t - z} \right),
\label{3}
\end{eqnarray}
where $k = \left( {\omega ,{\bf{k}}} \right)$ is the wave vector, $\delta  =  \pm 1$ is the ellipticity parameter of the wave and ${e_{x,y}} = \left( {0,{{\bf{e}}_{x,y}}} \right)$ are the polarization four-vectors of the wave, particularly $e_{x,y}^2 =  - 1,{\rm{ }}\left( {{e_{x,y}}k} \right) = {k^2} = 0$. We treat interaction with the Coulomb potential of the nucleus within the first Born approximation, therefore, we restrict ourselves with the condition $Z\alpha /v \ll 1$ ($Z$ is the nuclear charge, $\alpha$ is the fine structure constant).

\begin{figure}[h]
\includegraphics[width=0.5\linewidth]{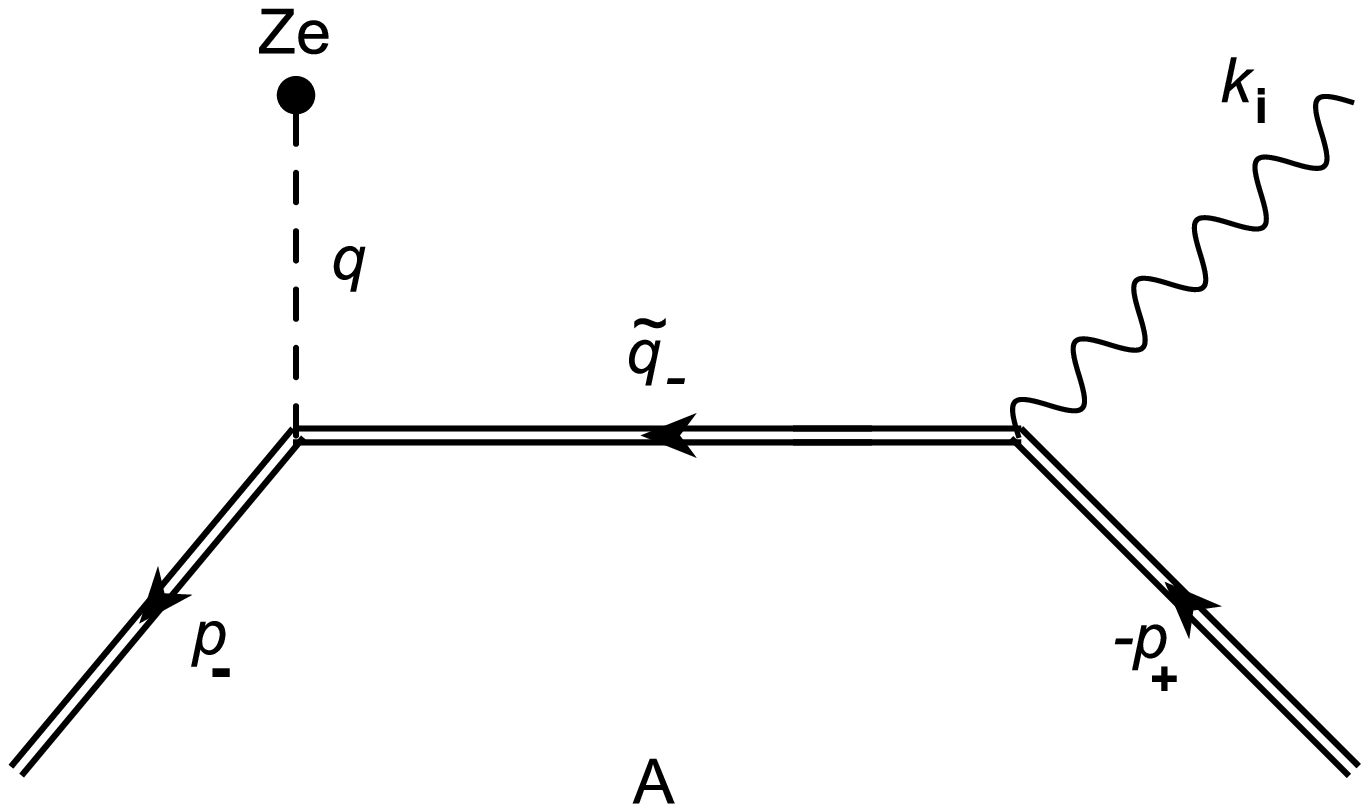}%
\hfill
\includegraphics[width=0.5\linewidth]{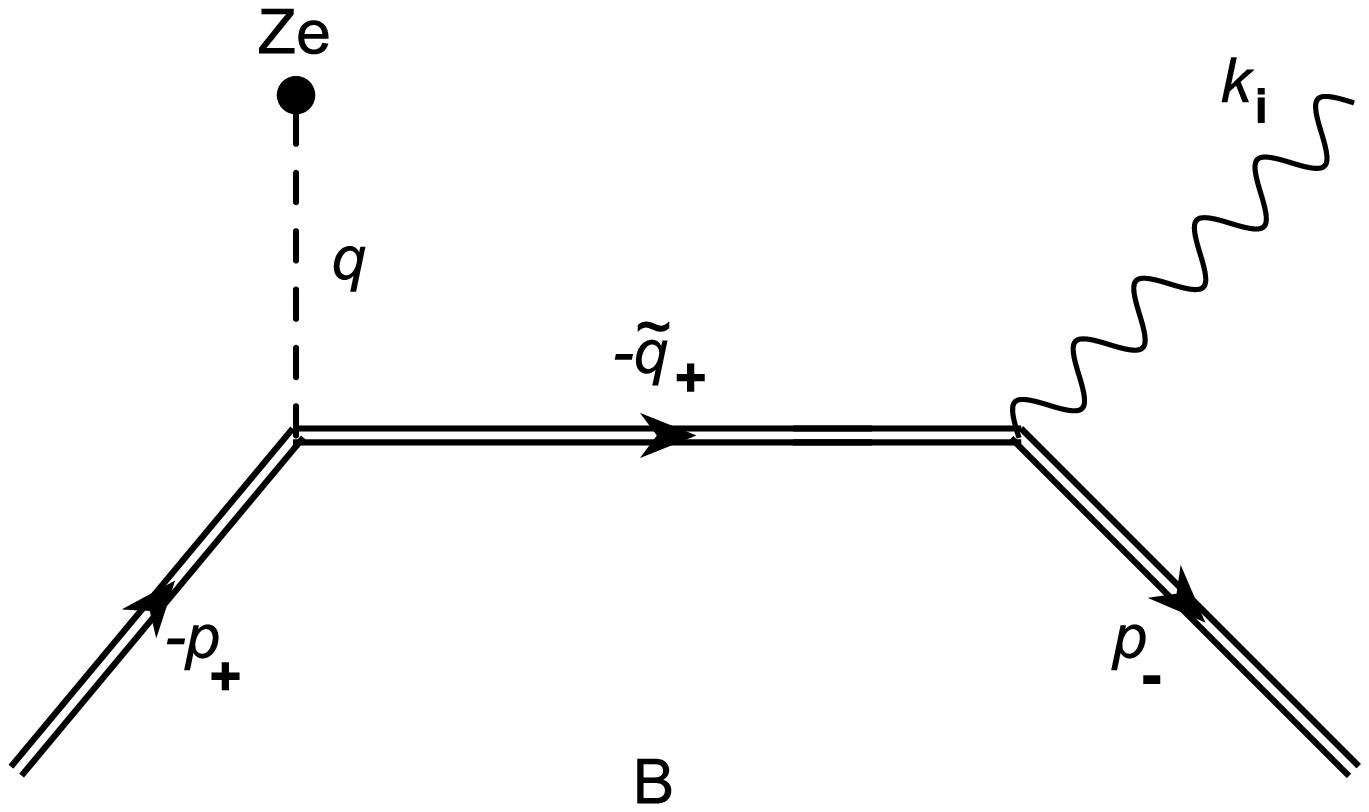}%
\caption{Feynman diagrams of the PPP process on the nucleus in the external electromagnetic field. Double incoming and outgoing lines correspond to the dressed electron and positron functions in the initial and final states. The inner lines stand for the intermediate dressed state, initial gamma quantum and pseudo photon of recoil are depicted by wavy and dashed lines accordingly.\label{Fig1}}%
\end{figure}

The considered process is of the second order in the fine structure constant, consequently it is described by two Feynman diagrams (see. Fig \ref{Fig1}), which differ from each other by the intermediate state.

The amplitude of such process can be represented as the sum over total number  absorbed (emitted) photons of external wave:
\begin{eqnarray}
S = \sum\limits_{l =  - \infty }^\infty  {{S_l},}
\label{4}
\end{eqnarray}
\begin{eqnarray}
{S_l} = \frac{{8Z{e^3}{\pi ^{5/2}}}}{{\sqrt {2{{\tilde E}_ + }{{\tilde E}_ - }{\omega _i}} }}\exp \left( {i\psi } \right)\left[ {{{\bar u}_{{p_ - },\lambda }}{B_l}{v_{{p_ + },\lambda '}}} \right]\frac{{\delta \left( {{q^0}} \right)}}{{{{\bf{q}}^2}}},
\label{5}
\end{eqnarray}
\begin{eqnarray}
{B_l} = \sum\limits_{r =  - \infty }^{ + \infty } {\left[ {{M_{r - l}}\left( {{{\tilde p}_ - },{{\tilde q}_ - }} \right)\frac{{{{\hat {\tilde q}}_ - } - \frac{{{\eta ^2}{m^2}}}{{2\left( {k{q_ - }} \right)}}\hat k + m}}{{\tilde q_ - ^2 - m_*^2}}{F_{ - r}}\left( {{{\tilde q}_ - },{{\tilde p}_ + }} \right) + {M_{r - l}}\left( {{{\tilde p}_ + },{{\tilde q}_ + }} \right)\frac{{{{\hat {\tilde q}}_ + } - \frac{{{\eta ^2}{m^2}}}{{2\left( {k{q_ + }} \right)}}\hat k + m}}{{\tilde q_ + ^2 - m_*^2}}{F_{ - r}}\left( {{{\tilde q}_ + },{{\tilde p}_ - }} \right)} \right]} 
\label{6}
\end{eqnarray}
Hereinafter, all notations with hat imply the contraction of the corresponding vector with Dirac gamma matrices ${\tilde \gamma ^\mu } = \left( {{{\tilde \gamma }^0},{\bf{\tilde \gamma }}} \right),{\rm{ }}\mu  = 0,1,2,3$ (e.g. $\hat k = {k_\mu }{\tilde \gamma ^\mu } = {k_0}{\tilde \gamma ^0} - {\bf{k\tilde \gamma }}$). In the expression (\ref{5}) ${\bar u_{{p_ - },\lambda }}$ and ${v_{{p_ + },\lambda '}}$ are free Dirac bispinors for electron in the final and positron in the initial state, respectively, and  $\psi$ is an insignificant phase that does not depend either on the summation index or momenta of particles. Here we introduced the notations for the electron and positron four-quasimomenta ${\tilde p_ \pm } = \left( {{{\tilde E}_ \pm },{{{\bf{\tilde p}}}_ \pm }} \right)$, as well four-quasimomenta of intermediate states ${\tilde q_ \pm } = \left( {{{\tilde E}_ \pm },{{{\bf{\tilde q}}}_ \pm }} \right)$:
\begin{eqnarray}
{\tilde q_ - } =  - {\tilde p_ + } + {k_i} + rk, \quad {\tilde q_ + } =  - {\tilde p_ - } + {k_i} + rk,
\label{7}
\end{eqnarray}
\begin{eqnarray}
q = {\tilde p_ + } + {\tilde p_ - } - {k_i} - lk.
\label{8}
\end{eqnarray}
\begin{eqnarray}
{\tilde p_ \pm } = {p_ \pm } + {\eta ^2}\frac{{{m^2}}}{{2\left( {k{p_ \pm }} \right)}}k,\quad {\tilde q_ \pm } = {q_ \pm } + {\eta ^2}\frac{{{m^2}}}{{2\left( {k{q_ \pm }} \right)}}k,
\label{9}
\end{eqnarray}
\begin{eqnarray}
\tilde p_ \pm ^2 = m_ * ^2,\quad {m_ * } = m\sqrt {1 + {\eta ^2}}.
\label{10}
\end{eqnarray}
Herein ${k_i} = {\omega _i}\left( {1,{{\bf{n}}_i}} \right)$ is the four-momentum of the initial gamma quantum and ${m_*}$ is an effective mass of fermion within the external electromagnetic field (\ref{3}).  The amplitudes ${M_{r - l}}$ and ${F_{ - r}}$ in the relation (\ref{6}) have the following expressions:
\begin{eqnarray}
{M_{l - r}}\left( {{{\tilde p}_2},{{\tilde p}_1}} \right) = {a^0}{L_{r - l}}\left( {{{\tilde p}_2},{{\tilde p}_1}} \right) + b_ - ^0{L_{r - l - 1}} + b_ + ^0{L_{r - l + 1}},\quad
\label{11}
\end{eqnarray}
\begin{eqnarray}
{F_{ - r}}\left( {{{\tilde p}_2},{{\tilde p}_1}} \right) = \left( {a\varepsilon } \right){L_{ - r}}\left( {{{\tilde p}_2},{{\tilde p}_1}} \right) + \left( {{b_ - }\varepsilon } \right){L_{ - r - 1}} + \left( {{b_ + }\varepsilon } \right){L_{ - r + 1}},
\label{12}
\end{eqnarray}
where we denoted by parentheses the dot product of the initial gamma quantum polarization four-vector  ${\varepsilon ^\mu }$ with matrices ${a^\mu },{\rm{ }}b_ \pm ^\mu $ that are defined in the following way:
\begin{eqnarray}
{a^\mu } = {\tilde \gamma ^\mu } + {\eta ^2}\frac{{{m^2}}}{{2\left( {k{{\tilde p}_1}} \right)\left( {k{{\tilde p}_2}} \right)}}{k^\mu }\hat k,
\label{13}
\end{eqnarray}
\begin{eqnarray}
b_ \pm ^\mu  = \frac{1}{4}\eta m\left[ {\frac{{{{\hat \varepsilon }_ \pm }\hat k{\gamma ^\mu }}}{{\left( {k{{\tilde p}_2}} \right)}} + \frac{{{\gamma ^\mu }\hat k{{\hat \varepsilon }_ \pm }}}{{\left( {k{{\tilde p}_1}} \right)}}} \right],{\rm{   }}{\hat \varepsilon _ \pm } = {\hat e_x} \pm i\delta {\hat e_y},
\label{14}
\end{eqnarray}
Special functions ${L_{r - l}}\left( {{{\tilde p}_2},{{\tilde p}_1}} \right)$,  ${L_{ - r}}\left( {{{\tilde p}_2},{{\tilde p}_1}} \right)$ and their arguments are given by the expressions \cite{L_fun}: 
\begin{eqnarray}
{L_n}\left( {{{\tilde p}_2},{{\tilde p}_1}} \right) = \exp \left( { - in{\chi _{{{\tilde p}_2}{{\tilde p}_1}}}} \right){J_n}\left( {{\gamma _{{{\tilde p}_2}{{\tilde p}_1}}}} \right)
\label{15}
\end{eqnarray}
\begin{eqnarray}
\tan {\chi _{{{\tilde p}_2}{{\tilde p}_1}}} = \delta \frac{{\left( {{e_y}{Q_{{{\tilde p}_2}{{\tilde p}_1}}}} \right)}}{{\left( {{e_x}{Q_{{{\tilde p}_2}{{\tilde p}_1}}}} \right)}},\quad {Q_{{{\tilde p}_2}{{\tilde p}_1}}} = \frac{{{{\tilde p}_2}}}{{\left( {k{{\tilde p}_2}} \right)}} + \frac{{{{\tilde p}_1}}}{{\left( {k{{\tilde p}_1}} \right)}},
\label{16}
\end{eqnarray}
\begin{eqnarray}
{\gamma _{{{\tilde p}_2}{{\tilde p}_1}}} = \eta m\sqrt { - Q_{{{\tilde p}_2}{{\tilde p}_1}}^2} .
\label{17}
\end{eqnarray}
We note, that $\left( {k{{\tilde p}_{1,2}}} \right) = \left( {k{p_{1,2}}} \right)$ and thus, to obtain the appropriate expressions for the channel A (i.e. for the first term in (\ref{6})) we need to replace ${\tilde p_1} \to  - {\tilde p_ + },{\rm{ }}{\tilde p_2} \to {\tilde q_ - }$ for ${F_{ - r}}\left( {{{\tilde p}_2},{{\tilde p}_1}} \right)$ and ${\tilde p_1} \to {\tilde q_ - },{\rm{ }}{\tilde p_2} \to {\tilde p_ - }$ for ${M_{l - r}}\left( {{{\tilde p}_2},{{\tilde p}_1}} \right)$ in the relations (\ref{13})-(\ref{17}). For the channel B (i.e. for the second term in (\ref{6})) one has to act in similar way and make the replacement ${\tilde p_1} \to {\tilde p_ - },{\rm{ }}{\tilde p_2} \to  - {\tilde q_ + }$ for ${F_{ - r}}\left( {{{\tilde p}_2},{{\tilde p}_1}} \right)$ and ${\tilde p_1} \to  - {\tilde q_ + },{\rm{ }}{\tilde p_2} \to  - {\tilde p_ + }$ for ${M_{l - r}}\left( {{{\tilde p}_2},{{\tilde p}_1}} \right)$. It is important to emphasize, that obtained amplitude (\ref{5})-(\ref{17}) is valid for the arbitrary intensities and frequencies of the plane monochromatic wave with circular polarization.

\section{POLES OF THE AMPLITUDE IN A STRONG FIELD}
\begin{figure}[h]
\includegraphics[width=0.5\linewidth]{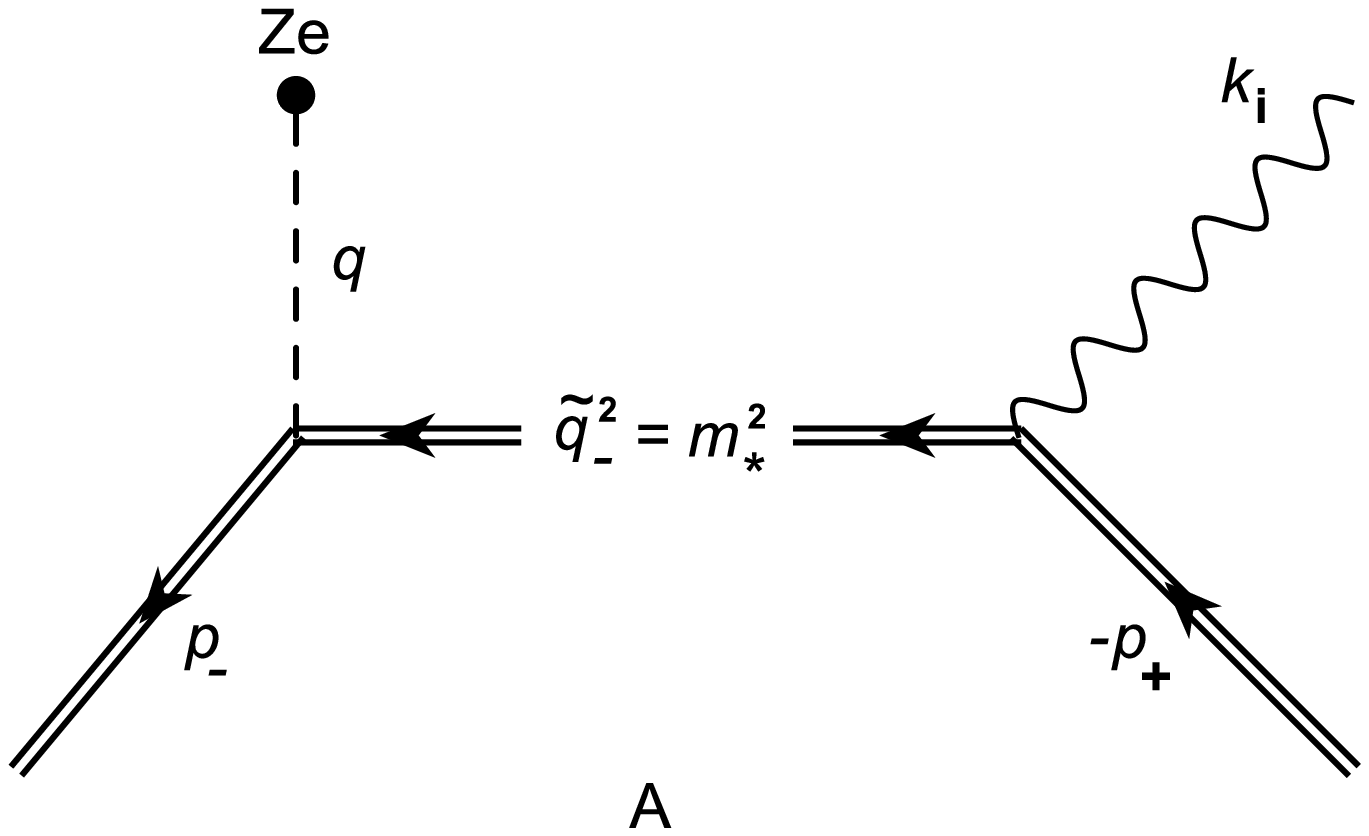}%
\hfill
\includegraphics[width=0.5\linewidth]{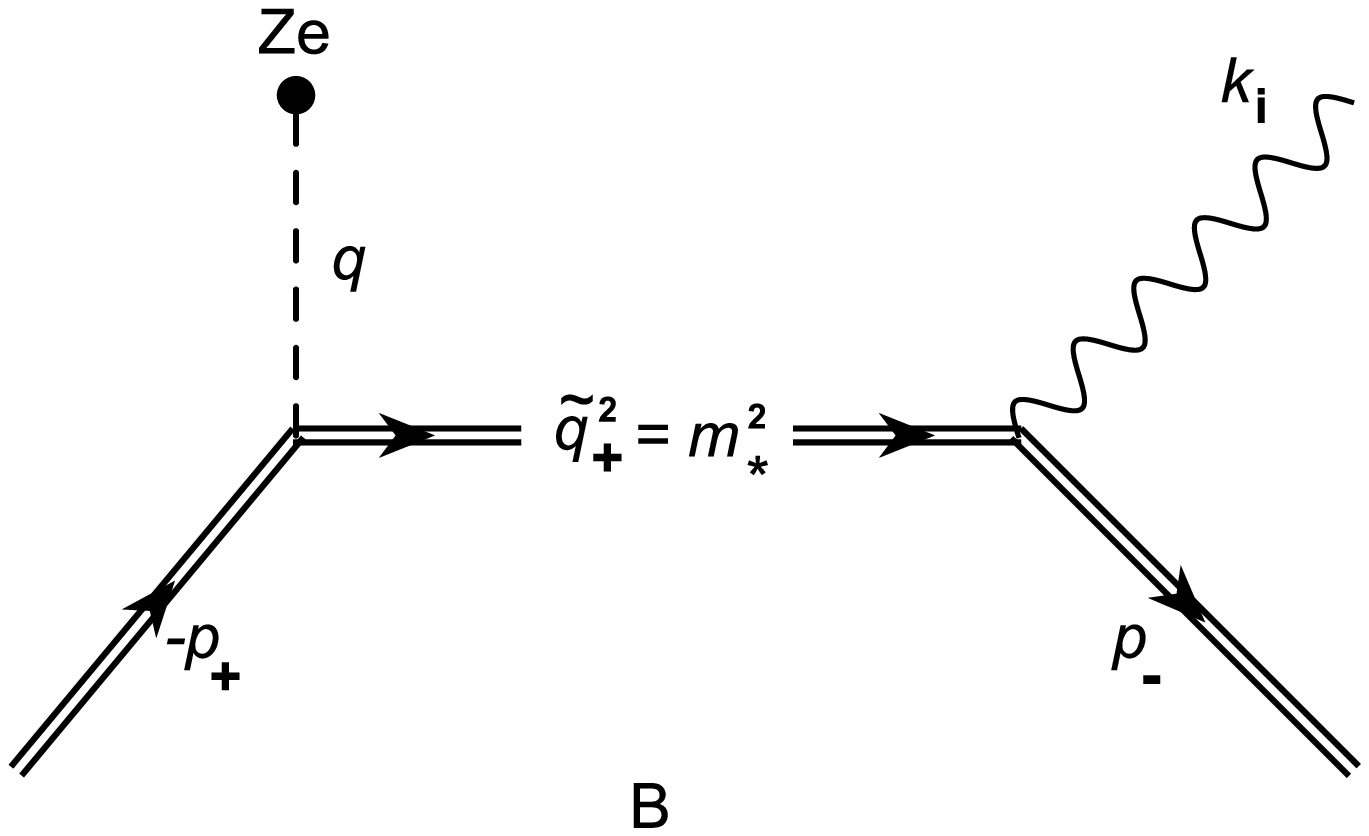}%
\caption{Resonant photoproduction of the electron-positron pair in the field of a nucleus and plane electromagnetic wave.\label{Fig2}}%
\end{figure}

In the presence of the external electromagnetic field (3) the intermediate particle momentum may satisfy its dispersion relation:
\begin{eqnarray}
\tilde q_ - ^2 = m_ * ^2,
\label{18}
\end{eqnarray}
\begin{eqnarray}
\tilde q_ + ^2 = m_ * ^2.
\label{19}
\end{eqnarray}
Such behavior is caused by the quasi-discrete energy spectrum of fermion propagating within the plane electromagnetic wave. Due to that fact, one may interpret it as the reduction of the second order process (see. Fig.\ref{Fig1}) into the two successive second order processes in fine structure constant (see Fig.\ref{Fig2}).

Within the external field of a plane monochromatic wave there is four-quasimomentum conservation law, which can be written for both channels in every vertex (see Fig.\ref{Fig2}) in the following way:
\begin{eqnarray}
{k_i} + rk = {\tilde p_ + } + {\tilde q_ - },
\label{20}
\end{eqnarray}
\begin{eqnarray}
q = {\tilde q_ - } - {\tilde p_ - } + \left( {l - r} \right)k
\label{21}
\end{eqnarray}
and
\begin{eqnarray}
{k_i} + rk = {\tilde p_ - } + {\tilde q_ + },
\label{22}
\end{eqnarray}
\begin{eqnarray}
q = {\tilde q_ + } - {\tilde p_ + } + \left( {l - r} \right)k.
\label{23}
\end{eqnarray}
Insofar as $\tilde p_ \pm ^2 = \tilde q_ \pm ^2 = m_ * ^2,{\rm{  and   }}\;{k^2} = k_i^2 = 0,$ the equalities (\ref{20}) and (\ref{22}) are fulfilled only for the $r \ge 1$. This fact in conjunction with the form of amplitude (\ref{5}), (\ref{6}) (see also Fig.\ref{Fig2}) allows us to conclude that ${F_{ - r}}$ (\ref{12}) represents the amplitude of the laser-stimulated Breit-Wheeler process \cite{Ritus} with absorption of $r$ wave photons. In turn, ${M_{r - l}}$ (\ref{11}) is nothing but laser-assisted Mott scattering of electron (channel A) or positron (channel B) on a nucleus with the absorption (emission) of $\left| {r - l} \right|$ wave photons \cite{Cajiao_V_lez_2019,Ehlotzky_rev}. Hence, that verified that in the absence of interference, the initial second order process in the fine structure constant effectively reduces into two successive first order processes, as was mentioned above. One also may check that simultaneous fulfillment of the resonant conditions (\ref{18}), (\ref{19}) and four-quasimomentum law conservation (\ref{20}), (\ref{22}) is impossible  unless the initial gamma quantum and external plane wave propagate in one direction.

The thorough examination of the resonant conditions (\ref{18}), (\ref{19}) and conservation laws (\ref{20})-(\ref{23}) shows us that for resonance to occur, one of the possibilities is to require the ultrarelativistic energies of produced particles and thus the sufficient energy of the initial gamma quantum. Moreover, the resonant kinematics region is confined with the configuration, where all produced particles have to propagate within the narrow cone with initial gamma quantum direction. Additionally, we demand the directions of initial gamma quantum and external wave propagation do not coincidence, otherwise resonances are merely impossible:
\begin{eqnarray}
{\omega _i} \gg m
\label{24}
\end{eqnarray}
\begin{eqnarray}
{\theta _{i \pm }} = \left( {{{\bf{k}}_i},{{\bf{p}}_ \pm }} \right) \ll 1,\quad {{\bar \theta }_ \pm } = \left( {{{\bf{p}}_ - },{{\bf{p}}_ + }} \right) \ll 1, \nonumber \\
{\theta _i} = \left( {{{\bf{k}}_i},{\bf{k}}} \right)\sim1,\quad {\theta _ \pm } = \left( {{\bf{k}},{{\bf{p}}_ \pm }} \right)\sim1.\quad
\label{25}
\end{eqnarray}
In a matter of fact, the condition (\ref{24}) has to be rewritten for the case of strong field, when the classical parameter becomes not small $\eta  \mathbin{\lower.3ex\hbox{$\buildrel>\over
{\smash{\scriptstyle\sim}\vphantom{_x}}$}} 1$. We face the necessity to replace the particle mass with the effective mass \cite{LarinSpie}. Hence, the condition (\ref{24}) takes form:
\begin{eqnarray}
\frac{{{\omega _i}}}{{{m_ * }}} = \frac{{{\omega _i}}}{{m\sqrt {1 + {\eta ^2}} }} \sim \left\{ {\begin{array}{*{20}{c}}
{{{{\omega _i}} \mathord{\left/
 {\vphantom {{{\omega _i}} {m \gg 1,\quad {\rm{if}}\quad \eta  \ll {\rm{1}}}}} \right.
 \kern-\nulldelimiterspace} {m \gg 1,\quad {\rm{if}}\quad \eta  \ll {\rm{1}}}}}\\
{{{{\omega _i}} \mathord{\left/
 {\vphantom {{{\omega _i}} {\left( {\eta m} \right) \gg 1,\quad {\rm{if}}\quad \eta  \mathbin{\lower.3ex\hbox{$\buildrel>\over
{\smash{\scriptstyle\sim}\vphantom{_x}}$}} {\rm{1}}}}} \right.
 \kern-\nulldelimiterspace} {\left( {\eta m} \right) \gg 1,\quad {\rm{if}}\quad \eta  \mathbin{\lower.3ex\hbox{$\buildrel>\over
{\smash{\scriptstyle\sim}\vphantom{_x}}$}} {\rm{1}}}}}
\end{array}} \right.
\label{26}
\end{eqnarray}
From the second string in the condition (\ref{26}) we obtain the restriction on the maximum intensity of the external field:
\begin{eqnarray}
\eta  \ll {\eta _{\max }} = \frac{{{\omega _i}}}{m}.
\label{27}
\end{eqnarray}
By the similar reasoning, we formulate the new ultrarelativistic condition for produced particles:
\begin{eqnarray}
\frac{{{{\tilde E}_ \pm }}}{{{m_ * }}} \approx \frac{{{E_ \pm }}}{{m\sqrt {1 + {\eta ^2}} }} \sim \left\{ {\begin{array}{*{20}{c}}
{{{{E_ \pm }} \mathord{\left/
 {\vphantom {{{E_ \pm }} {m \gg 1,\quad {\rm{if}}\quad \eta  \ll {\rm{1}}}}} \right.
 \kern-\nulldelimiterspace} {m \gg 1,\quad {\rm{if}}\quad \eta  \ll {\rm{1}}}}}\\
{{{{E_ \pm }} \mathord{\left/
 {\vphantom {{{E_ \pm }} {\left( {\eta m} \right) \gg 1,\quad {\rm{if}}\quad \eta  \mathbin{\lower.3ex\hbox{$\buildrel>\over
{\smash{\scriptstyle\sim}\vphantom{_x}}$}} {\rm{1}}}}} \right.
 \kern-\nulldelimiterspace} {\left( {\eta m} \right) \gg 1,\quad {\rm{if}}\quad \eta  \mathbin{\lower.3ex\hbox{$\buildrel>\over
{\smash{\scriptstyle\sim}\vphantom{_x}}$}} {\rm{1}}}}}
\end{array}} \right.
\label{28}
\end{eqnarray}

Deliberately, throughout our research, we consider the initial gamma quantum energy ${\omega _i} \mathbin{\lower.3ex\hbox{$\buildrel<\over
{\smash{\scriptstyle\sim}\vphantom{_x}}$}}  100\;{\rm{GeV}}$. This value leads us to the estimation of the classical invariant parameter $\eta  <  < {\eta _{\max }} \sim {10^5}$, which corresponds to $F <  < {F_{\max }} \sim {10^{15}}\;{{\rm{V}} \mathord{\left/
 {\vphantom {{\rm{V}} {{\rm{cm}}}}} \right.
 \kern-\nulldelimiterspace} {{\rm{cm}}}}$ $\left( {I <  < {I_{\max }} \sim {{10}^{28}}\;{{\rm{W}} \mathord{\left/
 {\vphantom {{\rm{W}} {{\rm{c}}{{\rm{m}}^{\rm{2}}}}}} \right.
 \kern-\nulldelimiterspace} {{\rm{c}}{{\rm{m}}^{\rm{2}}}}}} \right)$ for the optical frequency range. Therefore, all further results are valid for sufficiently large intensity, however, they are still not applicable to the fields of the critical Schwinger limit ${F_{cr}} \approx 1.3 \cdot {10^{16}}\;{{\rm{V}} \mathord{\left/
 {\vphantom {{\rm{V}} {{\rm{cm}}}}} \right.
 \kern-\nulldelimiterspace} {{\rm{cm}}}}$. 
 
 With use of relations (\ref{18}) and (\ref{20}) we can derive the expression for the resonant positron energy in channel A:
\begin{eqnarray}
{x_{\eta  + (r)}} = \frac{{r \pm \sqrt {r\left( {r - {r_\eta }} \right) - r_\eta ^2\delta _{\eta  + }^2} }}{{2\left( {r + r_\eta ^{}\delta _{\eta  + }^2} \right)}}.
\label{29}
\end{eqnarray}
Analogously, relations (19) and (22) help us to deduce expression for the resonant electron energy for channel B:
\begin{eqnarray}
{x_{\eta  - (r)}} = \frac{{r \pm \sqrt {r\left( {r - {r_\eta }} \right) - r_\eta ^2\delta _{\eta  - }^2} }}{{2\left( {r + {r_\eta }\delta _{\eta  - }^2} \right)}}.
\label{30}
\end{eqnarray}
Here we introduced notations:
\begin{eqnarray}
{x_{\eta  \pm (r)}} = \frac{{{E_{\eta  \pm (r)}}}}{{{\omega _i}}}{\rm{, }}\;{r_\eta }{\rm{ = }}\frac{{m_*^2}}{{{\omega _i}\omega {{\sin }^2}\left( {{{{\theta _i}} \mathord{\left/
 {\vphantom {{{\theta _i}} 2}} \right.
 \kern-\nulldelimiterspace} 2}} \right)}}{\rm{, }}\;{\delta _{\eta  \pm }} = \frac{{{\omega _i}{\theta _{i \pm }}}}{{2{m_*}}}.
\label{31}
\end{eqnarray}
In formulae (\ref{29}) and (\ref{30}) $r$ is a number of resonance (namely, it is the number of wave photons absorbed within the laser-stimulated BW process), ${r_\eta }$ is a characteristic parameter that determines the minimal amount of wave photons that are required for the laser-stimulated BW process to happen: $r\ge r_{min}$, where
\begin{eqnarray}
{r_{\min }} = \left\lceil {{r_\eta }} \right\rceil .
\label{32}
\end{eqnarray}
Throughout this paper, we will use for assessments the certain set of parameters: ${\omega _i} = 50\;{\rm{GeV}}$, $\omega  = 1\;{\rm{eV}}$, ${\theta _i} = \pi $. For such set, it follows:
\begin{eqnarray}
{r_\eta } \approx 5.2\left( {1 + {\eta ^2}} \right).
\label{33}
\end{eqnarray}
One can see that in strong fields $\left( {\eta  >  > 1} \right)$, the resonant process involves large number of absorbed wave photons $\left( {{r_\eta } \approx {\eta ^2} \gg 1} \right)$. By the definition of ${r_\eta }$ it follows, as well, that the number of absorbed photons increases proportional to intensity $\left( {{r_\eta } \sim {\eta ^2} \sim I\;\left( {{\rm{Wc}}{{\rm{m}}^{{\rm{ - 2}}}}} \right)} \right)$. We underline, that all obtained expressions (\ref{29})-(\ref{31}) are in complete agreement with the weak field limit  $\eta  \ll 1$, particularly the parameter ${r_\eta }$ reduces to the threshold energy for initial gamma quantum \cite{Larin}.

\begin{figure}[h]
\includegraphics[width=0.49\linewidth]{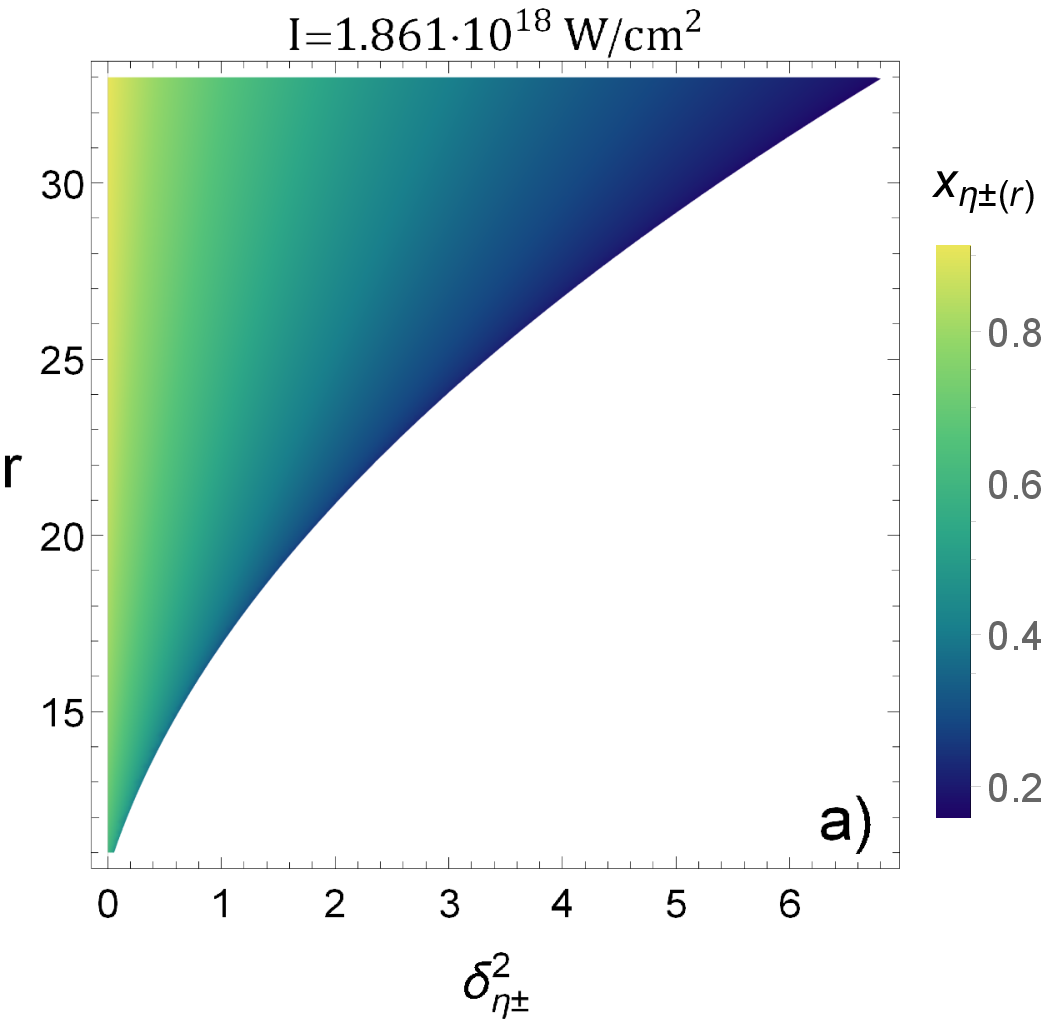}%
\hfill
\includegraphics[width=0.49\linewidth]{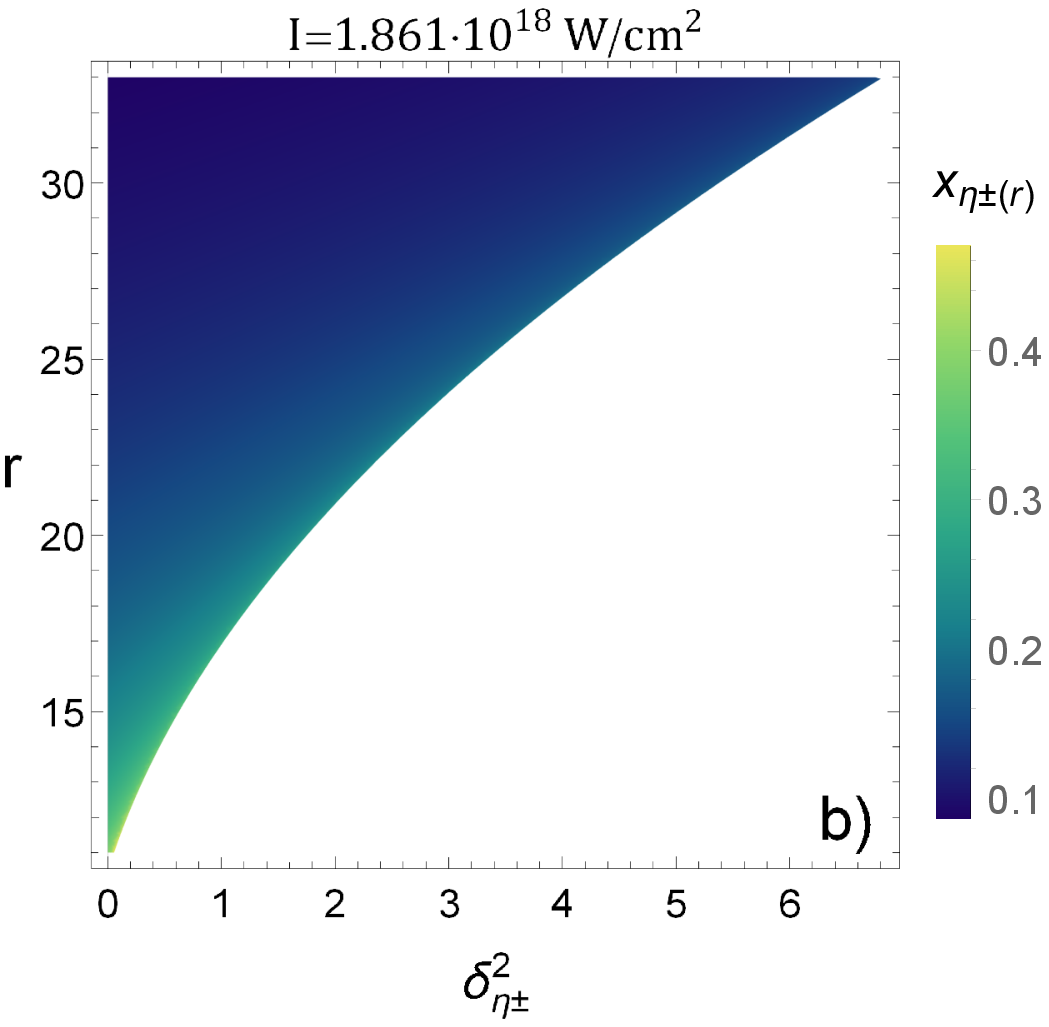}%
\vfill
\includegraphics[width=0.49\linewidth]{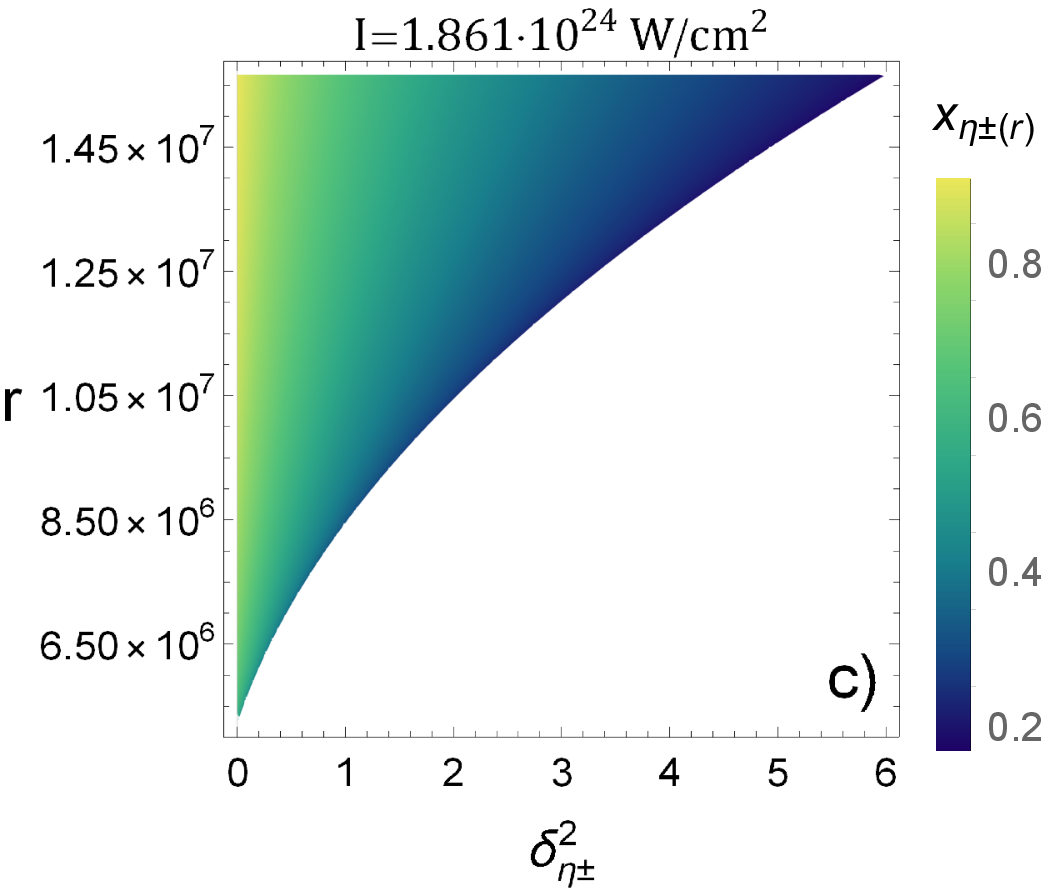}%
\hfill
\includegraphics[width=0.49\linewidth]{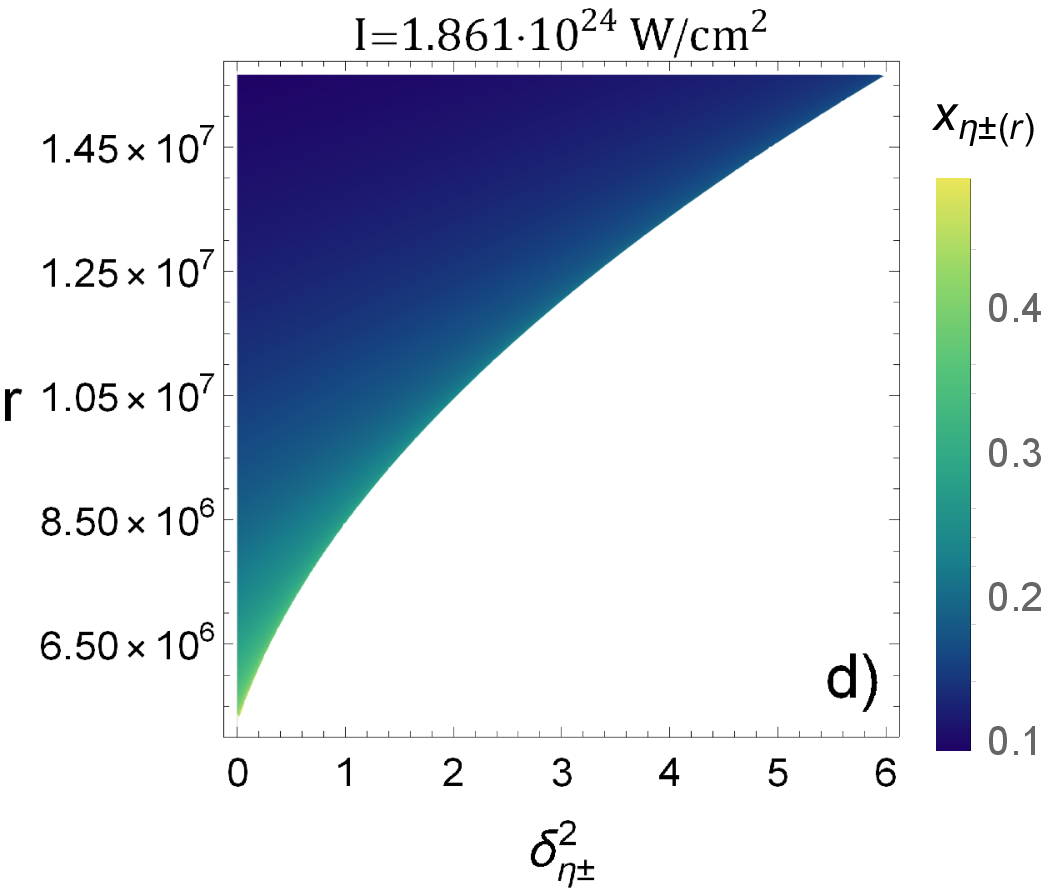}%
\caption{The resonant positron (channel A) and electron (channel B) energies as functions of absorbed wave photons and corresponding outgoing angle, plotted for the parameters (\ref{33}). Fig.3a and Fig.3c represent high-energy solution, meantime Fig.3b and Fig.3d correspond to the low-energy solutions (\ref{29}), (\ref{30}).\label{Fig3}}%
\end{figure}

Another peculiarity of the resonant behavior is that resonant energy of produced particles  ambiguously depends on corresponding outgoing angles (positron outgoing angle for channel A (\ref{29}) and electron outgoing angle for channel B (\ref{30})) (see. Fig.\ref{Fig3} and \cite{Larin,LarinLasPhys}). Henceforth, we will refer to the expressions with «+» sign in numerators of (\ref{29}) and (\ref{30}) as high-energy solutions and to expressions with «-» sign as low-energy. Also, the outgoing angle of particle (positron for channel A and electron for channel B) enclosed in the interval, which is defined by the following inequality:
\begin{eqnarray}
0 \le \delta _{\eta  \pm }^2 \le \delta _{\eta  \pm \max }^2 = \frac{r}{{{r_\eta }}}\left( {\frac{r}{{{r_\eta }}} - 1} \right).
\label{34}
\end{eqnarray}
\begin{figure}[h]
\includegraphics[width=0.5\linewidth]{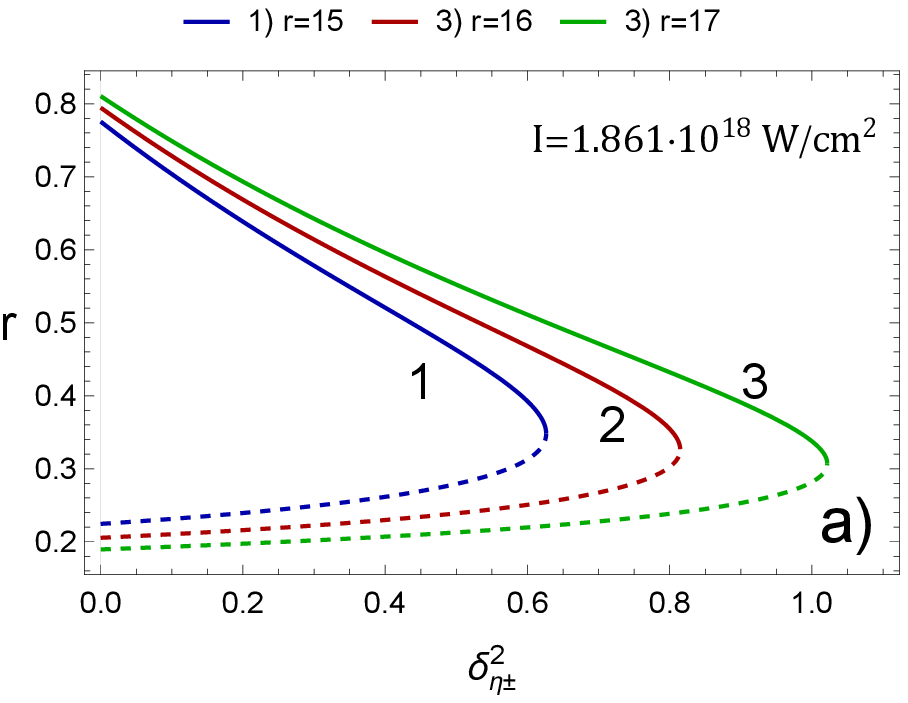}%
\hfill
\includegraphics[width=0.5\linewidth]{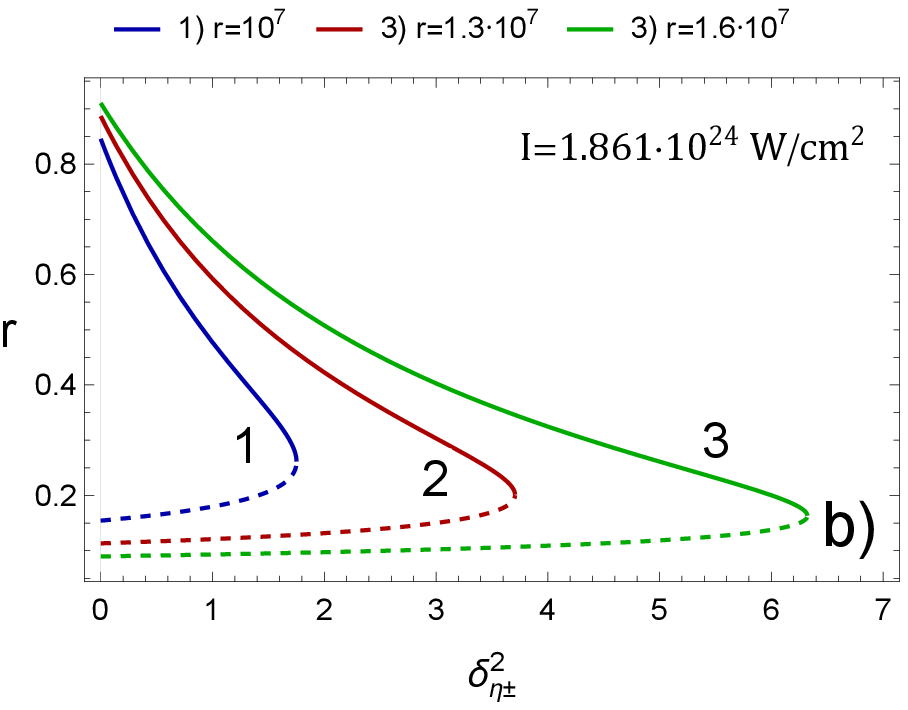}%
\caption{The resonant positron (channel A) and electron (channel B) energies as functions of the corresponding outgoing angle, plotted for the different values of absorbed wave photons and different intensities for certain values of the parameters (\ref{33}). Solid lines represent the high-energy solution, meantime dotted lines stand for the low-energy solutions (\ref{29}), (\ref{30}).\label{Fig4}}%
\end{figure}

Inasmuch there are no intersections between energy’s value with different   within the frame of particular channel (see. Fig.\ref{Fig4}), we can distinguish one process with different number of absorbed photons from another and thus, they do not interfere.

\section{THE RESONANT DIFFERENTIAL CROSS SECTION OF THE PPP IN THE ULTRARELATIVISTIC ENERGY LIMIT}

As long as we confine ourselves by the condition (\ref{27}) we may neglect the second and the third term in fermion scattering amplitude (\ref{11}) ($\left| {b_ \pm ^0} \right| \mathbin{\lower.3ex\hbox{$\buildrel<\over
{\smash{\scriptstyle\sim}\vphantom{_x}}$}} \eta {m \mathord{\left/
 {\vphantom {m {{\omega _i}}}} \right.
 \kern-\nulldelimiterspace} {{\omega _i}}} \ll 1$, see Eqs. (\ref{14}) and (\ref{26})). As a result, the expression for ${M_{r - l}}$ essentially simplifies:
\begin{eqnarray}
M_{r - l}^{} = \exp \left[ { - i\left( {r - l} \right){\chi _{{{\tilde p}_ - }{{\tilde q}_ - }}}} \right]{J_{r - l}}\left[ {\gamma \left( {{{\tilde p}_ - },{{\tilde q}_ - }} \right)} \right]{\gamma ^0}
\label{35}
\end{eqnarray}
For the conciseness, we represent our derivations for the channel A. To obtain the corresponding relations for channel B, one must substitute ${\tilde q_ - } \to {\tilde q_ + },\;{\tilde p_ - } \to  - {\tilde p_ + }$. Also, we introduce the subscript «+» for further expressions to signify that within the channel A all deduced quantities are defined by the positron outgoing angle, in contrast to the channel B, where we use the subscript «-» to accentuate the similar role of the electron outgoing angle. The influence of interference between different channels is left out of our consideration. Nevertheless, we emphasize that there is indeed interference in resonance case, and its impact requires further investigation. 

We perform the standard procedure \cite{LL} to derive resonant differential cross section for the unpolarized particles from the amplitude (\ref{4})-(\ref{6}), (\ref{12}), (\ref{35}):
\begin{eqnarray}
d\sigma _{ + (l,r)}^{} = \frac{2}{{{\pi ^2}}}d{{\rm M}_{ + \left( {l - r} \right)}}\frac{{{m^2}{E_ - }}}{{{{\left| {\tilde q_ - ^2 - m_ * ^2} \right|}^2}}}d{{\rm{P}}_{ + \left( r \right)}}.
\label{36}
\end{eqnarray}
Herein $d{{\rm M}_{ + \left( {l - r} \right)}}$ represents differential cross section of the intermediate electron scattering on the nucleus with emission (absorption) of $\left| {l - r} \right|$ wave photons (lasser-assisted Mott scattering) \cite{monmon}:
\begin{eqnarray}
d{{\rm M}_{ + \left( {l - r} \right)}} = {Z^2}r_e^2\frac{{{m^2}}}{{{{\bf{q}}^4}}}J_{l - r}^2\left( {{\gamma _{{{\tilde p}_ - },{{\tilde q}_ - }}}} \right)\delta \left[ {\tilde q_ - ^0 - {{\tilde E}_ - } + \left( {l - r} \right)\omega } \right]{d^3}{\tilde p_ - },\nonumber \\
\label{37}
\end{eqnarray}
where transferred to nucleus momentum ${\bf{q}}$ and argument of the Bessel function have the following forms: 
\begin{eqnarray}
{\bf{q}} = {{\bf{\tilde q}}_ - } - {{\bf{\tilde p}}_ - } + \left( {l - r} \right){\bf{k}},
\label{38}
\end{eqnarray}
\begin{eqnarray}
{\gamma _{{{\tilde p}_ - },{{\tilde q}_ - }}} = \eta m\sqrt { - Q_{{{\tilde p}_ - },{{\tilde q}_ - }}^2} ,\quad Q_{{{\tilde p}_ - },{{\tilde q}_ - }}^{} = \frac{{{{\tilde p}_ - }}}{{\left( {k{{\tilde p}_ - }} \right)}} - \frac{{{{\tilde q}_ - }}}{{\left( {k{{\tilde q}_ - }} \right)}}.\ 
\label{39}
\end{eqnarray}
Function $d{{\rm{P}}_{ + \left( r \right)}}$ determines the differential probability (per unit of time) of the lasser-stimulated BW process with the absorption of  wave photons \cite{Ritus}:
\begin{eqnarray}
d{{\rm{P}}_{ + \left( r \right)}} = \frac{\alpha }{{{\omega _i}{E_ + }}}P\left( {{u_{\eta  + \left( r \right)}},{v_{\eta \left( r \right)}}} \right){d^3}{\tilde p_ + },
\label{40}
\end{eqnarray}
here
\begin{eqnarray}
P\left( {{u_{\eta  + \left( r \right)}},{v_{\eta \left( r \right)}}} \right) = J_r^2\left( {{\gamma _{{{\tilde q}_ - }{{\tilde p}_ + }}}} \right) + {\eta ^2}\left( {2{u_{\eta  + \left( r \right)}} - 1} \right)\left[ {\left( {\frac{{{r^2}}}{{\gamma _{{{\tilde q}_ - }{{\tilde p}_ + }}^2}} - 1} \right)J_r^2 + J_r^{\prime 2}} \right],
\label{41}
\end{eqnarray}
\begin{eqnarray}
{\gamma _{\eta  + \left( r \right)}} = 2r\frac{\eta }{{\sqrt {1 + {\eta ^2}} }}\sqrt {\frac{{{u_{\eta  + \left( r \right)}}}}{{{v_{\eta \left( r \right)}}}}\left( {1 - \frac{{{u_{\eta  + \left( r \right)}}}}{{{v_{\eta \left( r \right)}}}}} \right)} ,
\label{42}
\end{eqnarray}
\begin{eqnarray}
{u_{\eta  + (r)}} = \frac{{{{\left( {k{k_i}} \right)}^2}}}{{4\left( {k{q_ - }} \right)\left( {k{p_ + }} \right)}} \approx \frac{1}{{4{x_{\eta  + \left( r \right)}}\left( {1 - {x_{\eta  + \left( r \right)}}} \right)}},
\label{43}
\end{eqnarray}
\begin{eqnarray}
{v_{\eta (r)}} = r\frac{{\left( {k{k_i}} \right)}}{{2m_ * ^2}} = \frac{r}{{{r_\eta }}}.
\label{44}
\end{eqnarray}
Due to the condition (\ref{28}) it is permissible to put ${d^3}{\tilde p_ - } \approx {d^3}{p_ - } \approx E_ - ^2d{E_ - }d{\Omega _ - }$ in the cross section (\ref{37}) and easily carry out the integration with respect to the electron energy (or the positron energy for channel B).

The appearance of the resonant infinity is caused by the idealized spatial and temporal dependence of the external filed, which allows analytical investigation. Complete treatment of such kind divergences involves cumbersome calculations of all radiative corrections to fermion propagator \cite{Oleinik1}. Other ways to elaborate this issue are to consider the finite size and duration of the external field \cite{LarinLasPhys,Mackenroth} or to engage the Sokhotski-Plemelj theorem\cite{Seipt,Ilderton}. The last method, in fact, also implies the finite duration of external laser pulse. In turn, we achieve the elimination of the resonant infinity in channels A and B by means of the renowned Breit-Wigner procedure \cite{BreitWigner}:
\begin{eqnarray}
{m_ * } \to {\mu _ * } = {m_ * } - i{\Gamma _{\eta  + \left( r \right)}},\quad {\Gamma _{\eta  + \left( r \right)}} = \frac{{\tilde q_ - ^0}}{{2{m_ * }}}W\left( {{r_\eta }} \right).
\label{45}
\end{eqnarray}
Here ${W_{\eta  + \left( r \right)}}$ is the total probability (per unit of time) of laser-stimulated BW process with absorption of  wave photons \cite{Ritus}:
\begin{eqnarray}
W\left( {{r_\eta }} \right) = \frac{{\alpha {m^2}}}{{8\pi {\omega _i}}}{\rm{P}}\left( {{r_\eta }} \right),
\label{46}
\end{eqnarray}
\begin{eqnarray}
{\rm{P}}\left( {{r_\eta }} \right) = \sum\limits_{n = {r_{\min }}}^\infty  {{{\rm{P}}_n}\left( {{r_\eta }} \right), } 
\label{47}
\end{eqnarray}
\begin{eqnarray}
{{\rm{P}}_n}\left( {{r_\eta }} \right) = \int\limits_1^{{n \mathord{\left/
 {\vphantom {n {{r_\eta }}}} \right.
 \kern-\nulldelimiterspace} {{r_\eta }}}} {\frac{{du}}{{u\sqrt {u\left( {u - 1} \right)} }}P\left( {u,\frac{n}{{{r_\eta }}}} \right)} ,
\label{48}
\end{eqnarray}
where function $P\left( {u,{n \mathord{\left/
 {\vphantom {n {{r_\eta }}}} \right.
 \kern-\nulldelimiterspace} {{r_\eta }}}} \right)$ is derived from the expressions (\ref{41})-(\ref{44}) by substitution: ${u_{\eta  + \left( r \right)}} \to u,\;r \to n$. Given the relations (\ref{45})-(\ref{48}) the resonant denominator for channel A can be represented as follows:
\begin{eqnarray}
{\left| {\tilde q_ - ^2 - \mu _ * ^2} \right|^2} = 16m_*^4x_{\eta  + \left( r \right)}^2\left[ {{{\left( {\delta _{\eta  + }^2 - \delta _{\eta  + \left( r \right)}^2} \right)}^2} + \Upsilon _{\eta  + \left( r \right)}^2} \right].\quad
\label{49}
\end{eqnarray}
Here $\Upsilon _{\eta  + \left( r \right)}^{}$ is the angular resonant width:
\begin{eqnarray}
{\Upsilon _{\eta  + \left( r \right)}} = \frac{\alpha }{{32\pi \left( {1 + {\eta ^2}} \right)}}\frac{{\left( {1 - {x_{\eta  + \left( r \right)}}} \right)}}{{{x_{\eta  + \left( r \right)}}}}{\rm{P}}\left( {{r_\eta }} \right).
\label{50}
\end{eqnarray}
In the Eq.(\ref{49}) parameter $\delta _{\eta  + \left( r \right)}^2$ is related to the positron resonant energy by the formula (\ref{29}), meanwhile parameter $\delta _{\eta  + }^2$ varies independently. We also note that the function ${\rm{P}}\left( {{r_\eta }} \right)$ has the most general form (\ref{47}) and valid for any values of the classical relativistic parameter $\eta$. However, for the case $\eta  \gg 1$ it is more convenient to work with another expression, that was deduced by Nikishov and Ritus \cite{Ritus}: 
\begin{eqnarray}
{\rm{P}}\left( \eta  \right) = \frac{3}{{16}}\sqrt {\frac{3}{2}} \frac{{\left( {1 + {\eta ^2}} \right)\eta }}{{{r_\eta }}}\exp \left[ { - \frac{4}{3}\frac{{{r_\eta }}}{{\left( {1 + {\eta ^2}} \right)\eta }}\left( {1 - \frac{1}{{15{\eta ^2}}}} \right)} \right],
\label{51}
\end{eqnarray}
In contrast to (\ref{47}) and (\ref{48}) in formula (\ref{51}) were performed summation over all processes with different $r$ and integration with respect to $u$. The corresponding calculations may be analytically carried out only for the case $\eta  \gg 1$. We emphasize that angular width ${\Upsilon _{\eta  + \left( r \right)}}$ increases with the wave intensity growth. Noteworthy that due to the field strength limitation, which we consider throughout this paper $F \mathbin{\lower.3ex\hbox{$\buildrel<\over
{\smash{\scriptstyle\sim}\vphantom{_x}}$}} {10^{14}}\;{{\rm{V}} \mathord{\left/
 {\vphantom {{\rm{V}} {{\rm{cm}}}}} \right.
 \kern-\nulldelimiterspace} {{\rm{cm}}}}$ (see Eq. (\ref{27})), the angular radiation width ${\Upsilon _{\eta  + \left( r \right)}}$ is significantly greater than the radiative corrections \cite{HartinWidth}. 
 
Further calculations with use of relations (\ref{26})-(\ref{30}) and (\ref{49}) lead us to the following expressions for cross sections for channel A and B:
\begin{eqnarray}
\frac{{d{\sigma _{\eta  \pm \left( {l,r} \right)}}}}{{d{x_{\eta  \pm \left( r \right)}}d\delta _{\eta  \pm }^2}} = \frac{{\left( {{Z^2}\alpha r_e^2} \right)}}{{\pi {{\left( {1 + {\eta ^2}} \right)}^2}}}\frac{{J_{l - r}^2\left( {{\alpha _{\eta  \pm \left( r \right)}}} \right)}}{{g_ \pm ^4}}\frac{{{{\left( {1 - {x_{\eta  \pm \left( r \right)}}} \right)}^3}}}{{\left[ {{{\left( {\delta _{\eta  \pm }^2 - \delta _{\eta  \pm \left( r \right)}^2} \right)}^2} + \Upsilon _{\eta  + \left( r \right)}^2} \right]{x_{\eta  \pm \left( r \right)}}}}P\left( {{u_{\eta  \pm \left( r \right)}},\frac{r}{{{r_\eta }}}} \right)d\delta _{\eta  \mp }^2d\varphi , \nonumber\\
\label{52}
\end{eqnarray}
where
\begin{eqnarray}
g_ \pm ^2 = {g_{\eta 0}} + {\left( {\frac{{{m_*}}}{{2{\omega _i}}}} \right)^2}{g_{\eta  \pm (r,l)}}.
\label{53}
\end{eqnarray}
Herein $\phi$ is the angle between planes $\left( {{{\bf{k}}_i},{{\bf{p}}_ + }} \right)$ and $\left( {{{\bf{k}}_i},{{\bf{p}}_ - }} \right)$. Similarly to the channel A, the parameter $\delta _{\eta  - \left( r \right)}^2$ is expressed via the resonant energy of electron for channel B (\ref{30}) and $\delta _{\eta  - }^2$ varies independently. The relativistic invariant parameter ${u_{\eta  - \left( r \right)}}$ and the resonant width for channel B obey to relations:
\begin{eqnarray}
{u_{\eta  - (r)}} = \frac{{{{\left( {k{k_i}} \right)}^2}}}{{4\left( {k{q_ + }} \right)\left( {k{p_ - }} \right)}} \approx \frac{1}{{4{x_{\eta  - \left( r \right)}}\left( {1 - {x_{\eta  - \left( r \right)}}} \right)}},
\label{54}
\end{eqnarray}
\begin{eqnarray}
{\Upsilon _{\eta  - \left( r \right)}} = \frac{\alpha }{{32\pi \left( {1 + {\eta ^2}} \right)}}\frac{{1 - {x_{\eta  - \left( r \right)}}}}{{{x_{\eta  - \left( r \right)}}}}{\rm{P}}\left( {{r_\eta }} \right).
\label{55}
\end{eqnarray}
The function $P\left( {{u_{\eta  - \left( r \right)}},{r \mathord{\left/
 {\vphantom {r {{r_\eta }}}} \right.
 \kern-\nulldelimiterspace} {{r_\eta }}}} \right)$ has the likewise to channel A form (\ref{41}) except the substitution ${u_{\eta  + \left( r \right)}} \to {u_{\eta  - \left( r \right)}}$. The impact of transferred to nucleus momentum contains in the functions $g_ + ^2$ and $g_ - ^2$ (\ref{53}), where we took into account the influence of corrections proportional to ${{m_ * ^2} \mathord{\left/
 {\vphantom {{m_ * ^2} {\omega _i^2}}} \right.
 \kern-\nulldelimiterspace} {\omega _i^2}}$:
\begin{eqnarray}
{g_{\eta 0}} = \tilde \delta _{\eta  + }^2 + \tilde \delta _{\eta  - }^2 + 2\tilde \delta _{\eta  + }^{}\tilde \delta _{\eta  - }^{}\cos \varphi ,\;\;{\rm{ }}\tilde \delta _{\eta  \pm }^{} = 2{x_{\eta  \pm (r)}}\delta _{\eta  \pm }^{},\quad
\label{56}
\end{eqnarray}
\begin{eqnarray}
{g_{\eta  \pm (r,l)}} = g_{\eta  \pm (r,l)}^{(0)} + \frac{1}{{1 + {\eta ^2}}}g_{\eta  \pm (r,l)}^{(1)} + \frac{1}{{{{\left( {1 + {\eta ^2}} \right)}^2}}}g_{\eta  \pm (r,l)}^{(2)},\quad
\label{57}
\end{eqnarray}
\begin{eqnarray}
g_{\eta  \pm (r,l)}^{(0)} = \frac{{\tilde \delta _{\eta  \pm }^2}}{3}\frac{{x_{\eta  \pm (r)}^2{{\left( {1 - {x_{\eta  \pm (r)}}} \right)}^2} - {{\left( {1 - {x_{\eta  \pm (r)}}} \right)}^3} - x_{\eta  \pm (r)}^3}}{{x_{\eta  \pm (r)}^3{{\left( {1 - {x_{\eta  \pm (r)}}} \right)}^3}}}-\frac{{4{\beta _{\eta  \pm (l,r)}}\tilde \delta _{\eta  \pm }^2}}{{x_{\eta  \pm (r)}^{}\left( {1 - {x_{\eta  \pm (r)}}} \right)}}\qquad
\label{58}
\end{eqnarray}
\begin{eqnarray}
g_{\eta  \pm (r,l)}^{(1)} = 2\tilde \delta _{\eta  \pm }^2\frac{{x_{\eta  \pm (r)}^3\left( {2 - {x_{\eta  \pm (r)}}} \right) - {{\left( {1 - {x_{\eta  \pm (r)}}} \right)}^4}}}{{x_{\eta  \pm (r)}^3{{\left( {1 - {x_{\eta  \pm (r)}}} \right)}^3}}} +\frac{{4{\beta _{\eta  \pm (l,r)}}}}{{x_{\eta  \pm (r)}^{}\left( {1 - {x_{\eta  \pm (r)}}} \right)}}, \qquad
\label{59}
\end{eqnarray}
\begin{eqnarray}
g_{\eta  \pm (r,l)}^{(2)} = \frac{{{{\left( {1 - {x_{\eta  \pm (r)}}} \right)}^3} + x_{\eta  \pm (r)}^3 - {{\left( {1 - 2{x_{\eta  \pm (r)}}} \right)}^2}}}{{x_{\eta  \pm (r)}^3{{\left( {1 - {x_{\eta  \pm (r)}}} \right)}^3}}},
\label{60}
\end{eqnarray}
\begin{eqnarray}
{\beta _{\eta  \pm (l,r)}} = \frac{l}{{{r_\eta }}} - \frac{1}{4}\frac{{{\eta ^2}}}{{1 + {\eta ^2}}}\frac{1}{{{x_{\eta  \pm (r)}}\left( {1 - {x_{\eta  \pm (r)}}} \right)}}.
\label{61}
\end{eqnarray}
The arguments of the Bessel functions that define the processes of emission or absorption of $\left| {l - r} \right|$ wave photons during the scattering of intermediate fermion on the nucleus for channel A and B have the form: 
\begin{eqnarray}
{\alpha _{\eta  \pm \left( r \right)}} \approx 2{r_\eta }\frac{\eta }{{\sqrt {1 + {\eta ^2}} }}\sqrt {g_{\eta 0}^2} 
\label{62}
\end{eqnarray}
As we have already mentioned, the corrections of the order of   were introduced in transmitted momentum (\ref{53}). These corrections are of the great importance for the certain kinematic regions. Namely, they make the dominant contribution to the differential cross section under the conditions:
\begin{eqnarray}
\left| {\varphi  - \pi } \right| \mathbin{\lower.3ex\hbox{$\buildrel<\over
{\smash{\scriptstyle\sim}\vphantom{_x}}$}} \frac{{{m_ * }}}{{{\omega _i}}} \ll 1,\quad \left| {{{\tilde \delta }_{\eta  + }} - {{\tilde \delta }_{\eta  - }}} \right| \mathbin{\lower.3ex\hbox{$\buildrel<\over
{\smash{\scriptstyle\sim}\vphantom{_x}}$}} \frac{{{m_ * }}}{{{\omega _i}}} \ll 1.
\label{63}
\end{eqnarray}
Under such conditions, the function $g_{\eta 0}^{}$ tends to zero and consequently, there is a sharp maximum in the corresponding differential cross section. This notorious behavior of differential cross section in ultrarelativistic limit is due to the long-range Coulomb potential \cite{LL}. We perform the saddle point method to integrate resonant cross sections (\ref{52}) (for the channel B analogically) within the vicinity of maxima points (\ref{63}):  
\begin{eqnarray}
\frac{{d{\sigma _{\eta  + \left( {l,r} \right)}}}}{{d{x_{\eta  + \left( r \right)}}d\delta _{\eta  + }^2}} = \frac{{\left( {{Z^2}\alpha r_e^2} \right)}}{{4\pi {{\left( {1 + {\eta ^2}} \right)}^2}}}\frac{{\left( {1 - {x_{\eta  + \left( r \right)}}} \right)P\left( {{u_{\eta  + \left( r \right)}},{r \mathord{\left/
 {\vphantom {r {{r_\eta }}}} \right.
 \kern-\nulldelimiterspace} {{r_\eta }}}} \right)}}{{\left[ {{{\left( {\delta _{\eta  + }^2 - \delta _{\eta  + \left( r \right)}^2} \right)}^2} + \Upsilon _{\eta  + \left( r \right)}^2} \right]{x_{\eta  + \left( r \right)}}}}{C_{\eta  + \left( {l,r} \right)}},
\label{64}
\end{eqnarray}
where 
\begin{eqnarray}
{C_{\eta  + \left( {l,r} \right)}} = \int\limits_0^{2\pi } {d\varphi } \int\limits_0^\infty  {J_{l - r}^2\left( {{\alpha _{\eta  + \left( r \right)}}} \right){{\left[ {g_{\eta 0}^2 + {{\left( {\frac{{{m_*}}}{{2{\omega _i}}}} \right)}^2}{g_{\eta  + (r,l)}}} \right]}^{ - 2}}} d\tilde \delta _{\eta  - }^2.
\label{65}
\end{eqnarray} 
The integrand in (\ref{65}) has an abrupt maximum within the interval (\ref{63}) herewith, the function $J_{l - r}^2\left( {{\alpha _{\eta  + \left( r \right)}}} \right) \mathbin{\lower.3ex\hbox{$\buildrel<\over
{\smash{\scriptstyle\sim}\vphantom{_x}}$}} 1$ and we are allowed to take it out of the integral
\begin{eqnarray}
{C_{\eta  + \left( {l,r} \right)}} \approx J_{l - r}^2\left( {{\alpha _{\eta  + \left( r \right)}}} \right)\int\limits_0^{2\pi } {d\varphi } \int\limits_0^\infty  {d\tilde \delta _{\eta  - }^2} \exp \left[ {{f_\eta }\left( {\varphi ,{{\tilde \delta }_{\eta  - }}} \right)} \right],
\label{66}
\end{eqnarray}
here
\begin{eqnarray}
{f_\eta }\left( {\varphi ,{{\tilde \delta }_{\eta  - }}} \right) =  - 2\ln \left( {g_{\eta 0}^2 + {\kappa _{\eta  + }}} \right),\quad {\kappa _{\eta  + }} = {\left( {\frac{{{m_*}}}{{2{\omega _i}}}} \right)^2}{g_{\eta  + (r,l)}}.
\label{67}
\end{eqnarray}
We employ the Taylor expansion of the ${f_\eta }$ in the vicinity of the point $\varphi  = \pi ,\;\;{\tilde \delta _{\eta  - }} = {\tilde \delta _{\eta  + }}$
\begin{eqnarray}
{f_\eta }\left( {\varphi ,{{\tilde \delta }_{\eta  - }}} \right) \approx  - 2\ln \left( {{\kappa _{\eta  + }}} \right) - \frac{{2\tilde \delta _{\eta  + }^2}}{{{\kappa _{\eta  + }}}}{\left( {\varphi  - \pi } \right)^2} -\frac{1}{{2{\kappa _{\eta  + }}\tilde \delta _{\eta  + }^2}}{\left( {\tilde \delta _{\eta  - }^2 - \tilde \delta _{\eta  + }^2} \right)^2}.
\label{68}
\end{eqnarray}
Eventually, we perform integration (\ref{66}) with use of the expansion (\ref{68})
\begin{eqnarray}
{C_{\eta  + \left( {l,r} \right)}} \approx \frac{\pi }{{{g_{\eta  + \left( {l,r} \right)}}}}\left( {\frac{{2\omega _i^2}}{{m_ * ^2}}} \right)J_{l - r}^2\left( 0 \right) = \frac{{4\pi }}{{{g_{\eta  + \left( r \right)}}}}\left( {\frac{{\omega _i^2}}{{m_ * ^2}}} \right).
\label{69}
\end{eqnarray}
In so doing, we take into account that the argument of Bessel function (\ref{62}) is virtually zero and thus, the Bessel function itself is not zero only for the case $l=r$. In other words, the most probable situation is the scattering of ultrarelativistic fermion on the nucleus without absorption or emission of wave photons\cite{Lebed__2016,Larin}. Ultimately, the differential cross section for both channels may be written in the following way:
\begin{eqnarray}
\frac{{d{\sigma _{\eta  \pm \left( {l,r} \right)}}}}{{d{x_{\eta  \pm \left( r \right)}}d\delta _{\eta  \pm }^2}} = {\left( {\frac{{{\omega _i}}}{{{m_*}}}} \right)^2}\frac{{\left( {{Z^2}\alpha r_e^2} \right)}}{{{{\left( {1 + {\eta ^2}} \right)}^2}}}\frac{{\left( {1 - {x_{\eta  \pm \left( r \right)}}} \right)}}{{{g_{\eta  \pm \left( r \right)}}\left[ {{{\left( {\delta _{\eta  \pm }^2 - \delta _{\eta  \pm \left( r \right)}^2} \right)}^2} + \Upsilon _{\eta  \pm \left( r \right)}^2} \right]{x_{\eta  + \left( r \right)}}}}P\left( {{u_{\eta  \pm \left( r \right)}},\frac{r}{{{r_\eta }}}} \right), \nonumber \\
\label{70}
\end{eqnarray}
Herein the functions ${g_{\eta  + \left( r \right)}}$ and ${g_{\eta  - \left( r \right)}}$ are defined by the expressions (\ref{57})-(\ref{61}) but instead of the index $l$ one should write $r$. When the following conditions are met
\begin{eqnarray}
{\left( {\delta _{\eta  + }^2 - \delta _{\eta  + \left( r \right)}^2} \right)^2} \ll \Upsilon _{\eta  + \left( r \right)}^2,\: {\left( {\delta _{\eta  - }^2 - \delta _{\eta  - \left( r \right)}^2} \right)^2} \ll \Upsilon _{\eta  - \left( r \right)}^2\quad
\label{71}
\end{eqnarray}
we obtain the maximum resonant differential cross section for channels A and B:
\begin{eqnarray}
R_{\eta  \pm \left( r \right)}^{\max } = \frac{{d\sigma _{\eta  \pm \left( r \right)}^{\max }}}{{d{x_{\eta  \pm \left( r \right)}}d\delta _{\eta  \pm }^2}} = \left( {{Z^2}\alpha r_e^2} \right){F_{\eta  \pm \left( r \right)}}.
\label{72}
\end{eqnarray}
The functions ${F_{\eta  + \left( r \right)}}$ and ${F_{\eta  - \left( r \right)}}$ determine the spectral-angular distribution at fixed intensity of the resonant PPP differential cross section for channels A and B, correspondingly:
\begin{eqnarray}
{F_{\eta  \pm \left( r \right)}} = D\frac{{{x_{\eta  \pm \left( r \right)}}}}{{\left( {1 - {x_{\eta  \pm \left( r \right)}}} \right){g_{\eta  \pm \left( r \right)}}{{\rm{P}}^2}\left( {{r_\eta }} \right)}}P\left( {{u_{\eta  \pm \left( r \right)}},\frac{r}{{{r_\eta }}}} \right),
\label{73}
\end{eqnarray}
\begin{eqnarray}
D = {\left( {\frac{{32\pi {\omega _i}}}{{\alpha {m_*}}}} \right)^2}.
\label{74}
\end{eqnarray}

\begin{figure}[h]
\includegraphics[width=0.49\linewidth]{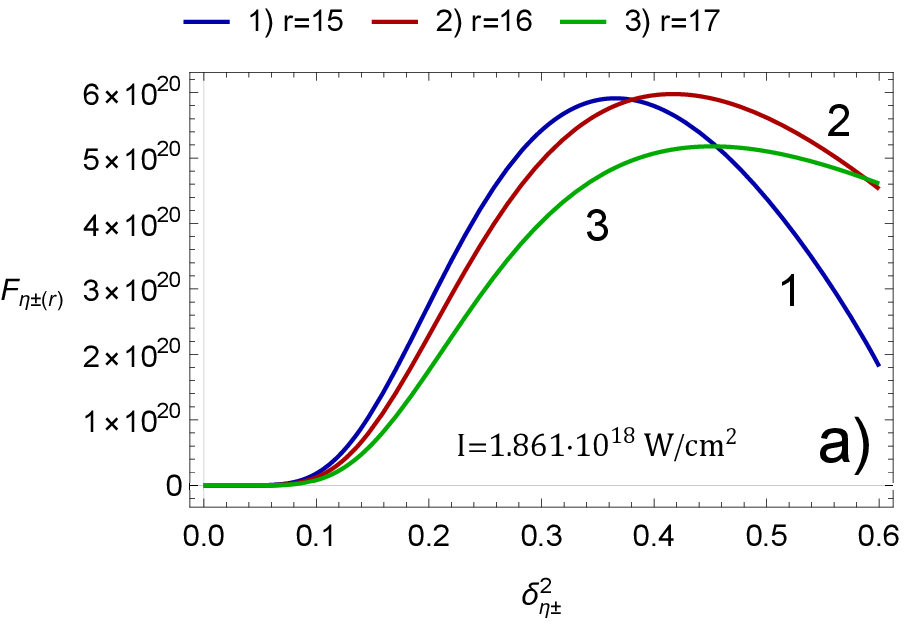}%
\hfill
\includegraphics[width=0.5\linewidth]{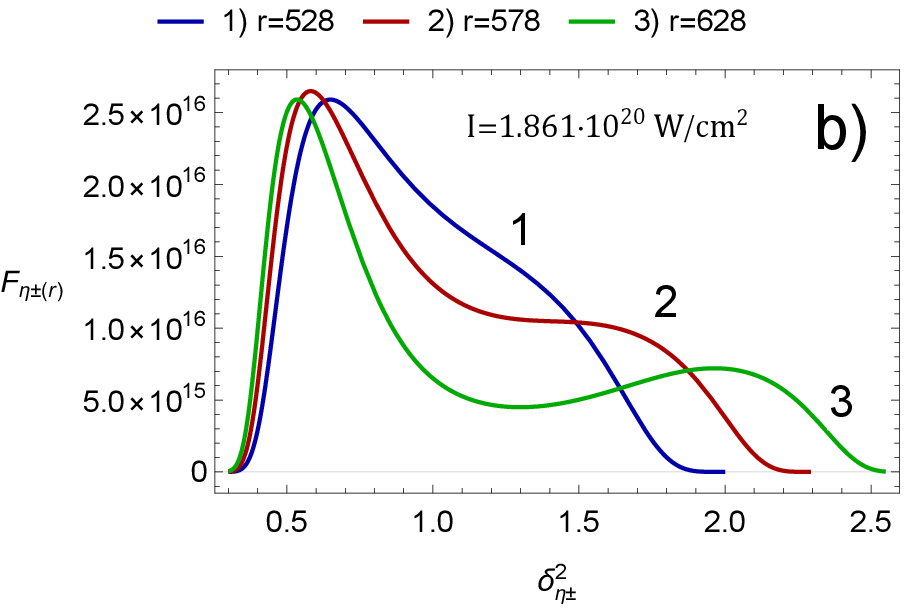}%
\vfill
\includegraphics[width=0.5\linewidth]{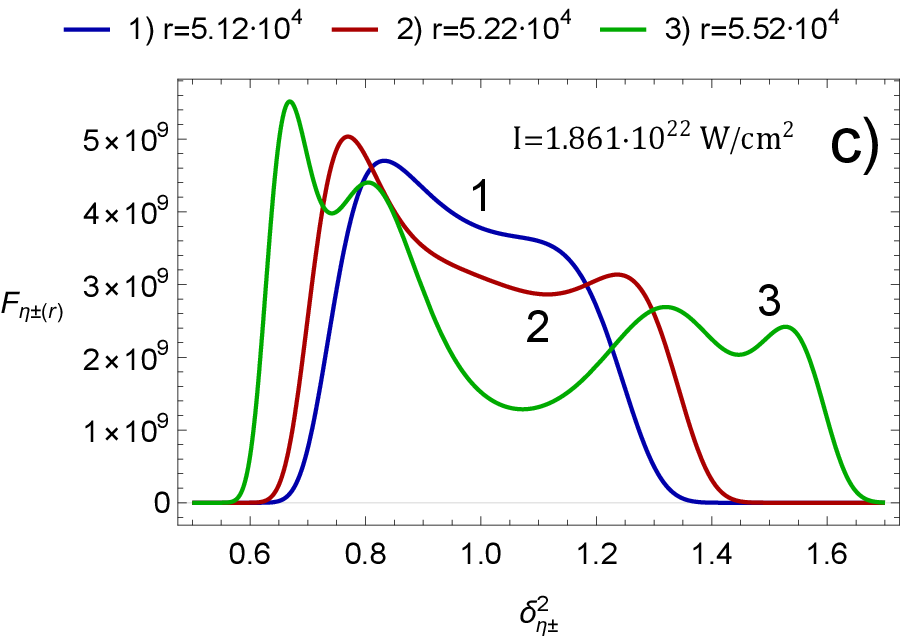}%
\hfill
\includegraphics[width=0.48\linewidth]{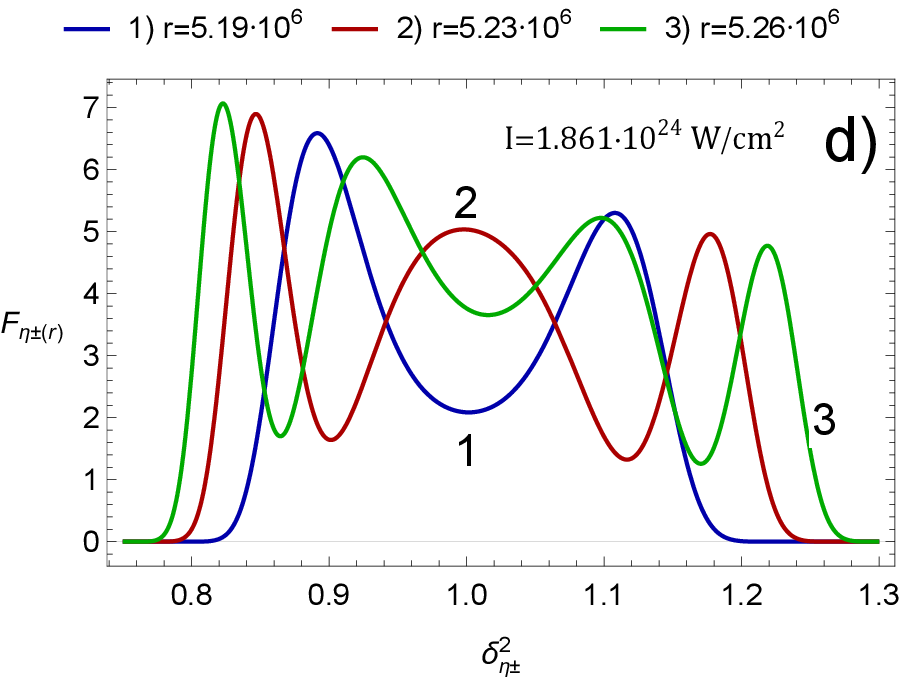}%
\caption{Dependence of the resonant cross section (\ref{73}) (in units ${Z^2}\alpha r_e^2$) on the corresponding outgoing angle for the certain parameters (\ref{33}) and different intensity values\label{Fig5}}%
\end{figure}

\begin{figure}[h]
\includegraphics[width=0.5\linewidth]{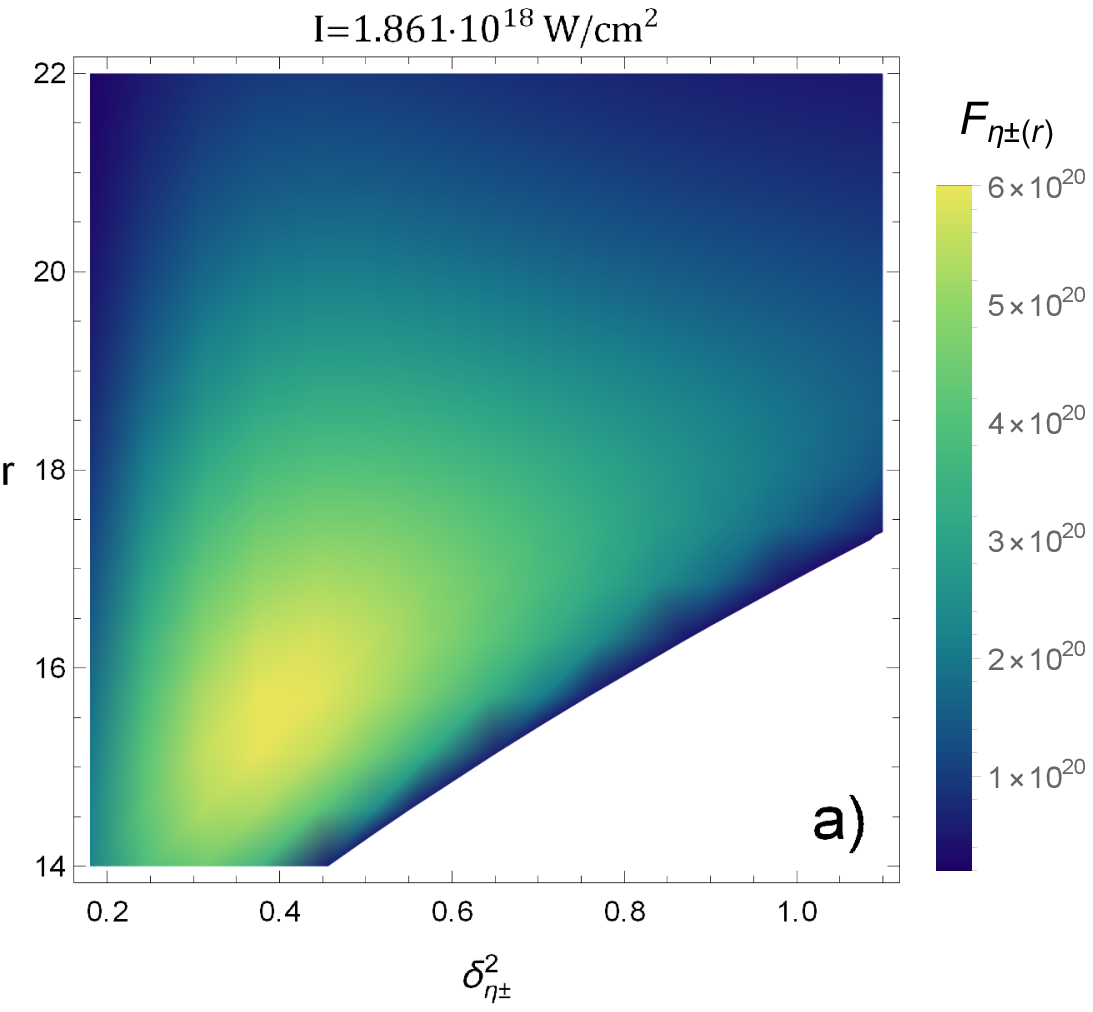}%
\hfill
\includegraphics[width=0.5\linewidth]{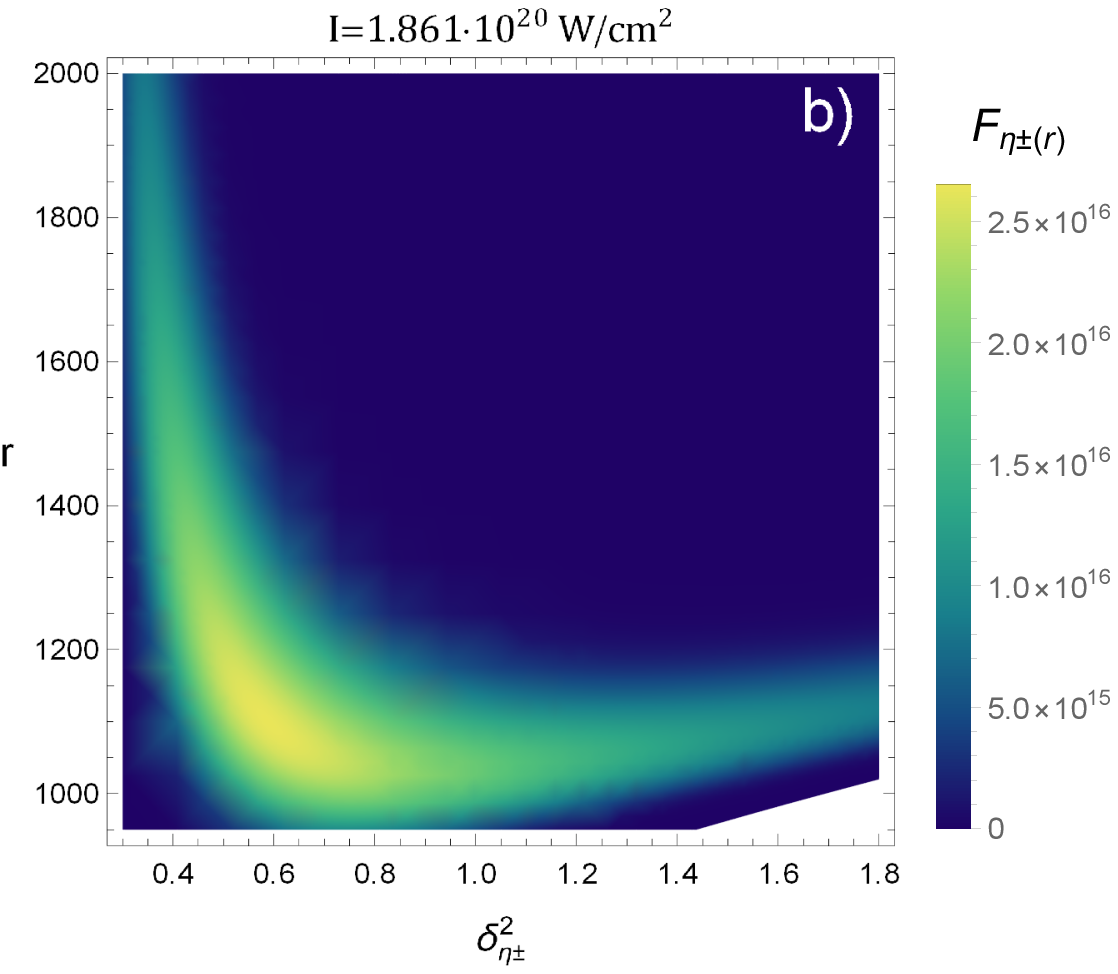}%
\vfill
\includegraphics[width=0.5\linewidth]{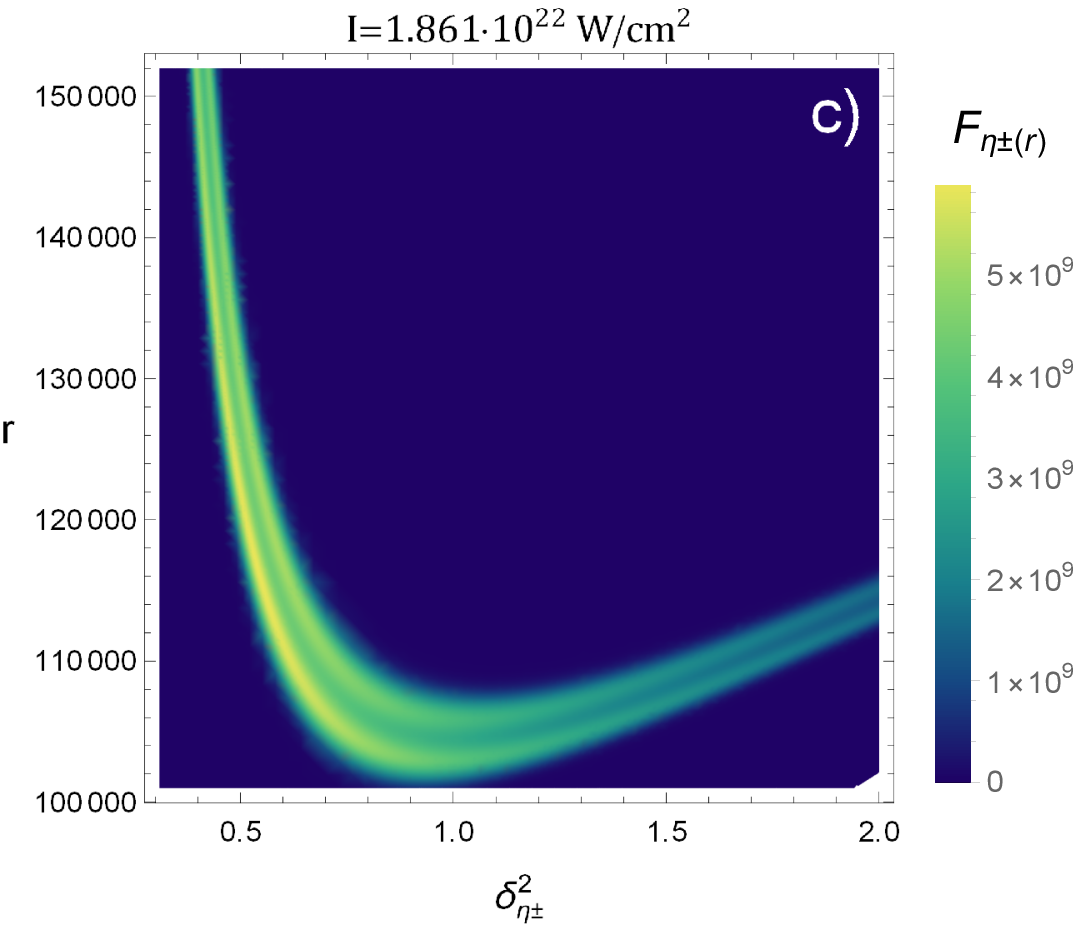}%
\hfill
\includegraphics[width=0.5\linewidth]{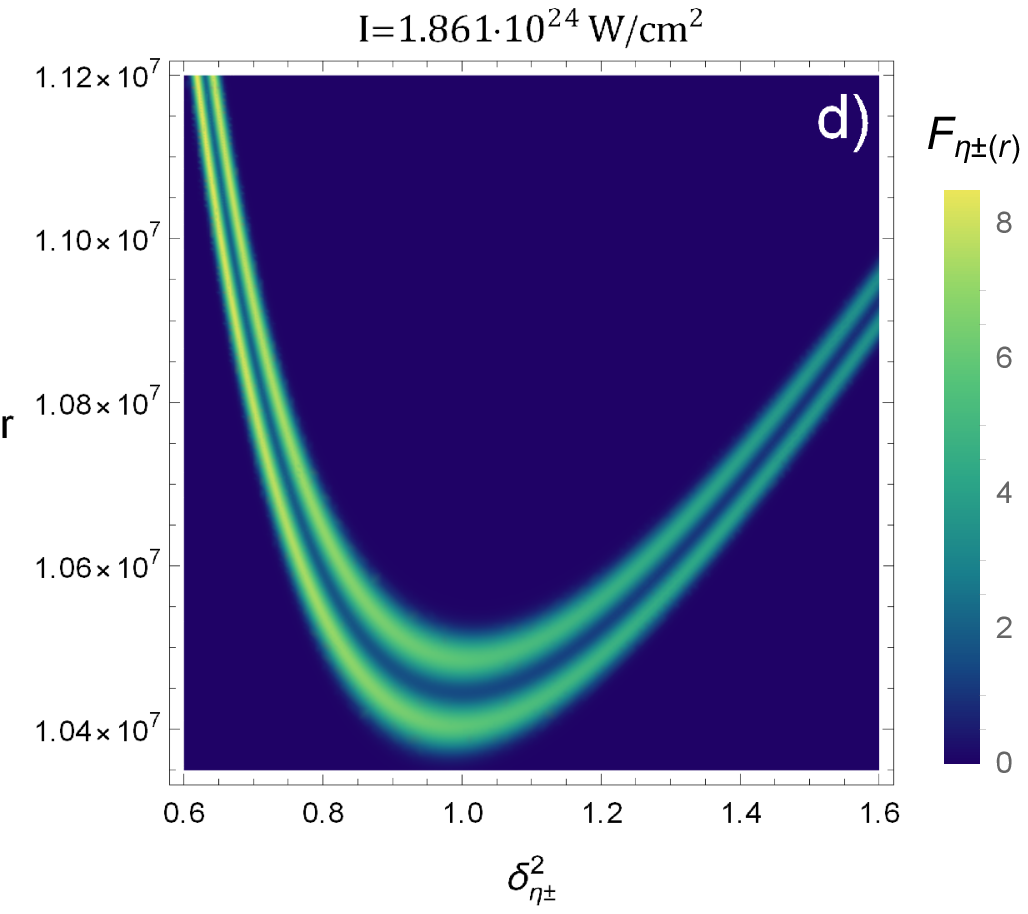}%
\caption{Temperature maps of the resonant cross section with respect to the corresponding outgoing angle and amount of absorbed wave photons for different intensities and parameters (\ref{33}) \label{Fig6}}%
\end{figure}

From Fig.\ref{Fig6}a-Fig.\ref{Fig6}d we can conclude, that only for particular values of the outgoing angles and absorbed wave photons resonant differential cross sections have significant magnitudes. In plane ($\delta_\pm^2,r$) these values form a region, which shape depends on the intensity. More detailed behavior of this region is presented in Fig.\ref{Fig5}a-Fig.\ref{Fig5}d, where we can see that with increase of intensity, the number of maxima grows, and they become more distinguishable. Herewith, the global maximum exists for every value of intensity, and therefore it allows us to determine the most probable energies of produced particles and corresponding outgoing angles. We stress, that the magnitude of the resonant differential cross section decreases with increase of intensity. Such dependence is explained by the behavior of the resonant angular width (\ref{50}), (\ref{55}), which increases with intensity. It is noteworthy, that we represented in the Fig.\ref{Fig5} and Fig.\ref{Fig6} the corresponding expressions with "high-energy" solutions (\ref{29}), (\ref{30}) therein. The reason is, that expressions with "low-energy" solutions utterly suppressed in comparison to the "high-energy" ones and don't make any impact into the total resonant cross section.

\section{Conclusion}
We have considered the resonant photoproduction of electron-positron pair on a nucleus within the strong external field. The thorough examination of four-quasimomentum conservation laws along with resonant conditions allowed us to formulate one of the possible resonant conditions. Due to these conditions, we have to demand ultrarelativistic energies of the produced particles and consequently sufficient energy of the initial gamma quantum. Besides, the propagation of the produced particles has to be enclosed within the narrow cone in the initial gamma quantum direction. We established, that there is minimal amount of absorbed wave photons in resonance. This amount is completely determined by the experimental set up and increases proportional to the external field intensity. The energies of produced particles were derived as the functions of outgoing angle and number of absorbed photons. The corresponding dependencies are ambiguous. In addition, there is maximum outgoing angle for every value of absorbed photons.

We obtained the resonant differential cross section of laser-assisted BH process with simultaneous registration of the particle energies and the corresponding outgoing angle (positron for channel A and electron for channel B) for vast intensity range (from $10^{18}$ to $10^{24}\;{\rm{W/c}}{{\rm{m}}^2}$). Also, we verified that under the resonant conditions,  it factorizes into the product of differential probability of BW process and differential cross section of Mott scattering. Herewith, the most probable is the situation, when intermediate particle scatters on nucleus without absorption or emission of wave photons. The obtained angular distribution has distinguishable maxima for each value of the absorbed wave photons. The number of maxima varies from one to four, depending on the number of absorbed photons and intensity. Moreover, there are global maxima of the resonant cross section with respect to the outgoing angle and number of absorbed wave photons for certain intensity value. This fact gives opportunity to determine the most probable energies of the particles and their outgoing angles, and thus to define a resonant process with high accuracy. Noteworthy, that with increase of the intensity the resonant differential cross section decreases due to the growth of the resonant width. There are also kinematic regions, where resonant differential cross section is totally suppressed.

The above analysis was carried out with considering the model of plane monochromatic electromagnetic wave, which allowed us to provide analytical investigation, but led us to notorious divergence of the resonant cross section. We employed the phenomenological procedure to eliminate this divergence, whereas, we emphasized that there are other more elegant and, in a matter of fact, more rigorous ways to deal with this problem. However, by virtue of inconsistency of the mentioned methods with the model of plane monochromatic wave, we can't perform them here. The detailed analysis involving proper elaboration of resonant infinity and influence of interference between different channels will be a subject of future research.

\section{Acknowledgment}
The work was supported by the State Program for the Fundamental Research (theme code FSEG-2020-0024)
 
\bibliography{MRE}

\end{document}